\documentclass[aps, prd, twocolumn, preprintnumbers, superscriptaddress, nofootinbib, floatfix]{revtex4-1}

\usepackage{amsmath}	
\usepackage{amsthm}	
\usepackage{amssymb}	
\usepackage{datetime}
\usepackage{graphicx}
\usepackage{verbatim}
\usepackage{bm}
\usepackage{soul}
\usepackage{hyperref}
\usepackage{epstopdf}
\usepackage{hyperref}

\enlargethispage{\baselineskip}

\hypersetup{backref,
colorlinks=true,
linkcolor=blue,
linktoc=page,
citecolor=blue,
urlcolor=blue}

\newcommand{\sv}{\langle\sigma v\rangle}
\newcommand{\rKIN}{\rho_{\rm kin}}

\newcommand{\xKIN}{x_{\rm kin}}

\newcommand{\TKIN}{T_{\rm kin}}
\newcommand{\Tfo}{T_{\rm f.o}}

\newcommand{\HH}{\mathcal{H}}

\newcommand{\MP}{M_{\rm Pl}}

\newcommand{\be}{\begin{equation}}
\newcommand{\ee}{\end{equation}}
\newcommand{\bea}{\begin{eqnarray}}
\newcommand{\eea}{\end{eqnarray}}
\renewcommand\({\left(}
\renewcommand\){\right)}
\renewcommand\[{\left[}
\renewcommand\]{\right]}

\begin{document}

\title{\bf (Non-)Thermal Production of WIMPs during Kination}

\newcommand{\FIRSTAFF}{\affiliation{Department of Physics and Astronomy,
	Uppsala University,
	L\"agerhyddsv\"agen 1,
	75120 Uppsala,
	Sweden}}
\newcommand{\SECONDAFF}{\affiliation{Nordita, KTH Royal Institute of Technology and Stockholm University,
	Roslagstullsbacken 23,\\
	10691 Stockholm,
	Sweden}}
	
\author{Luca Visinelli}
\email[Electronic address: ]{luca.visinelli@physics.uu.se}
\FIRSTAFF
\SECONDAFF

\date{\today}
\preprint{NORDITA-2017-114}

\begin{abstract}
Understanding the nature of the Dark Matter (DM) is one of the current challenges in modern astrophysics and cosmology. Knowing the properties of the DM particle would shed light on physics beyond the Standard Model and even provide us with details of the early Universe. In fact, the detection of such a relic would bring us information from the pre-Big Bang Nucleosynthesis (BBN) period, an epoch from which we have no direct data, and could even hint at inflation physics. In this work, we assume that the expansion rate of the Universe after inflation is governed by the kinetic energy of a scalar field $\phi$, in the so-called ``kination'' model. Adding to previous work on the subject, we assume that the $\phi$ field decays into both radiation and DM particles, which we take to be Weakly Interacting Massive Particles (WIMPs). The present abundance of WIMPs is then fixed during the kination period through either a thermal ``freeze-out'' or ``freeze-in'' mechanism, or through a non-thermal process governed by the decay of $\phi$. We explore the parameter space of this theory with the requirement that the present WIMP abundance provides the correct relic budget. Requiring that BBN occurs during the standard cosmological scenario sets a limit on the temperature at which the kination period ends. Using this limit and assuming the WIMP has a mass $m_\chi = 100\,$GeV, we obtain that the thermally-averaged WIMP annihilation cross section has to satisfy the constraints $4 \times 10^{-16}{\rm \,GeV^{-2}} \lesssim \langle\sigma v\rangle \lesssim 2\times 10^{-5} {\rm \,GeV^{-2}}$ in order for having at least one of the production mechanism to yield the observed amount of DM. This result shows how the properties of the WIMP particle, if ever measured, can yield information on the pre-BBN content of the Universe.
\end{abstract}

\maketitle

\section{Introduction}

The existence of a Dark Matter (DM) component in the Universe has long been established~\cite{Jungman:1995df, Bertone:2004pz}, with a Weakly Interacting Massive Particle (WIMP) being among the best motivated particle candidates~\cite{Ade:2015tva, Ade:2015xua}. In the simplest scenario of the early Universe, WIMPs of mass $m_\chi$ interact with the Standard Model (SM) particles at a sufficiently high rate so that the chemical equilibrium is attained. Owing to the expansion rate of the Universe, when the temperature falls below $T_{\rm f.o.} \approx m_\chi / 20$ WIMPs chemically decouple from the plasma and ``freeze-out'' of the equilibrium distribution~\cite{Vysotsky:1977pe, Hut:1977zn, Sato:1977ye, Lee:1977ua, Dicus:1977qy, Steigman:1979kw, Bernstein:1985th, Kolb:1985nn}. After freeze-out, the number of WIMPs in a comoving volume is fixed and the WIMP relic abundance is preserved to present day, assuming that there is no subsequent change in the entropy of the matter-radiation fluid. Coincidentally, the thermally-averaged WIMP annihilation cross section needed to explain the observed DM is of the same order of magnitude as that obtained for a process mediated by weakly interactions. For a WIMP of mass $m_\chi = 100\,$GeV, the annihilation cross section that provides the observed amount of DM satisfies $\sv_{\rm std} \approx 2\times 10^{-9}\,$GeV$^{-2}$. WIMPs continue to exchange momentum through elastic collisions with the plasma even after chemical decoupling, until this second mechanism also becomes inefficient and WIMPs decouple kinetically at a temperature $T_{\rm kd}$. Typically, $T_{\rm kd}$ ranges between $10\,$MeV and a few GeV~\cite{Profumo:2006bv}.

Even when considering this thermal production in the standard cosmological scenario, many caveats allow to alter the predicted relic density. Besides co-annihilation~\cite{Griest:1990kh, Edsjo:1997bg}, annihilation into forbidden channels~\cite{Griest:1990kh, DAgnolo:2015ujb, Cline:2017tka}, a momentum- or spin-dependent cross section~\cite{Chang:2009yt, Fan:2010gt, Fitzpatrick:2012ix}, or Sommerfeld enhancement~\cite{Hisano:2004ds, Lattanzi:2008qa, Cirelli:2008pk, ArkaniHamed:2008qn, Pospelov:2008jd, Fox:2008kb, Iengo:2009ni}, one possibility is that the annihilation cross section into the SM sector is so low that WIMPs never reach thermal equilibrium, effectively ``freezing-in'' to the present relic density, see for example Refs.~\cite{Hall:2009bx, Co:2015pka, Bernal:2017kxu}.

WIMPs might also be produced non-thermally, through the decay of a parent particle $\phi$~\cite{Kamionkowski:1990ni}, whose putative existence is motivated by the post-inflation reheating scenario. In facts, assume an early stage of inflation at an energy scale $H_I$ driven by one massive scalar field $\rho$ (the inflaton), of mass $m_\rho$. When slow-roll is violated, around $m_\rho \sim H_I$, inflation ends and the inflaton field reheats the Universe by decaying into lighter degrees of freedom~\cite{Kofman:1994rk, Kofman:1997yn}. We are not entering the details of the reheating mechanism here. In the standard picture, a radiation-dominated period begins as soon as the inflaton field has decayed and reheated the Universe.

In some reheating models the inflaton might also decay into one or more additional hypothetical fields, say for example a field $\phi$ of mass $m_\phi$ which comes to dominate the post-inflation stage. The inclusion of a non-standard cosmology between the reheating epoch right after inflation and the standard radiation-dominated scenario is motivated in realistic models of inflation, and might sensibly alter the WIMP relic density. A non-standard period might have lasted for a considerable amount of time, namely since the end of inflation down to a temperature which, using considerations on the Big Bang Nucleosynthesis (BBN) mechanism~\cite{Kawasaki:1999na, Kawasaki:2000en, Hannestad:2004px, Ichikawa:2005vw, DeBernardis:2008zz}, can be as low as $\sim 5\,$MeV. In these modified cosmologies, various properties of the WIMPs like their free-streaming velocity and the temperature at which the kinetic decoupling occurs have been investigated~\cite{Gelmini:2008sh, Visinelli:2015eka, Waldstein:2016blt, Waldstein:2017wps}.

In the Low Reheat Temperature Scenario (LRTS)~\cite{Dine:1982ah, Steinhardt:1983ia, Turner:1983he, Scherrer:1984fd}, $\phi$ is a massive modulus which drives an early matter-dominated epoch, eventually decaying into Standard Model particles and, possibly, WIMPs. In the pre-BBN LRTS, the thermal (both freeze-out and freeze-in) and non-thermal production of WIMPs have both been extensively studied~\cite{Lyth:1995ka, Chung:1998rq, Giudice:2000ex, Moroi:2000, Fujii:2002kr, Fujii:2003iw, Gelmini:2006pw, Gelmini:2006pq, Gelmini:2008sh, Acharya:2009zt, Grin:2010zz, Harigaya:2014waa, Baer:2014eja, Monteux:2015qqa, Reece:2015lch, Kane:2015qea, Erickcek:2015jza, Kim:2016spf}.

In the Kination Scenario (KS)~\cite{Barrow:1982, Ford:1986sy, Spokoiny:1993kt, Joyce:1996cp, Salati:2002md, Profumo:2003hq}, $\phi$ is a ``fast-rolling'' field whose kinetic energy governs the expansion rate of the post-inflation Universe, with an equation of state relating the pressure $p_\phi$ and energy density $\rho_\phi$ of the fluid as $p_\phi = \rho_\phi$. Owing to the scaling of the energy density in radiation with the scale factor $\rho_R \sim a^{-4}$, which is slower than the scaling of the energy density in the $\phi$ field $\rho_\phi \sim a^{-6}$, the contribution from the radiation energy density in determining the expansion rate eventually becomes more important than that from the $\phi$ field. When the $\phi$ field redshifts away, the standard radiation-dominated cosmology takes place. Since the scalar field $\phi$ dominates the expansion rate for some period, KS differs from the superWIMP model of Refs.~\cite{Feng:2003uy, Feng:2004zu}. Thermal production of WIMPs in the KS has been discussed in Refs.~\cite{Salati:2002md, Pallis:2005hm, Gomez:2008qi, Lola:2009at, Lewicki:2016efe, Artymowski:2016tme}, and has recently been investigated in Ref.~\cite{Redmond:2017tja}, in light of recent data. Ref.~\cite{DEramo:2017gpl} discussed the ``relentless'' thermal freeze-out in models where the $\phi$ field has a pressure $p_\phi > \rho_\phi/3$, thus including KS as an important sub-case. Ref.~\cite{Pallis:2005bb} discusses an intermediate model between LRTS and KS, in which a sub-dominant massive scalar field reheats the Universe during a kination period governed by an additional field.

In this paper, in addition to the thermal mechanisms of production recently discussed in Refs.~\cite{DEramo:2017gpl, Redmond:2017tja}, we assume that the $\phi$ field driving kination might decay into both radiation and WIMPs, with a decay rate $\Gamma_\phi$ and a branching ratio into WIMPs equal to $b$. Contrarily to previous KS models, kination ends when $\Gamma_\phi$ is equal to the expansion rate of the Universe, so when the $\phi$ field has decayed instead of being redshifted away. WIMP production proceeds by assuming that, before the Universe gets to be dominated by radiation at a temperature $\TKIN \gtrsim 5\,$MeV, the KS occurs. In the following, the subscript ``kin'' labels a quantity evaluated at $\TKIN$. We consider the thermal freeze-out and freeze-in mechanisms of WIMP production, governed by the thermal-averaged annihilation cross section times velocity $\sv$. We also include the non-thermal WIMP production from the decay of the $\phi$ field. We check that the WIMP population is always under-abundant with respect to other forms of energy. In summary, we show that in the model the present WIMP relic abundance can be reached through four different methods, namely the thermal ($b = 0$) or non-thermal ($b \neq 0$) production, either with or without ever reaching chemical equilibrium, as occurs in the LRTS~\cite{Gelmini:2006pw, Gelmini:2006pq}.

\section{Boltzmann equations for the model}

Energy conservation in an expanding homogeneous and isotropic universe is expressed as
\be
	\dot\rho_T + 3H(p_T+\rho_T) = 0,
\ee
where $\rho_T$ and $p_T$ are the total energy density and pressure of the contents, and $H$ is the Hubble expansion rate. We express the total energy density and pressure as a sum of three components, respectively describing the contents in the $\phi$ field, radiation, and the WIMPs. We model the evolution of these components by a set of coupled Boltzmann equations~\cite{Lyth:1995ka, Chung:1998rq, Giudice:2000ex, Moroi:2000}
\begin{eqnarray}
	\dot{\rho}_\phi + 3H(p_\phi + \rho_\phi) \!&=&\! -\Gamma_\phi \rho_\phi, \label{eq:boltzmann_phi}\\
	\dot{\rho}_R \!+\! 3H(p_R + \rho_R) \!&=& (1-b)\Gamma_\phi \rho_\phi + \frac{\sv}{\epsilon_\chi} \left(\rho_\chi^2 - \rho_{\rm EQ}^2\right)\!\!, \label{eq:boltzmann_R} \\
	\dot{\rho}_\chi + 3H\,(1+w)\,\rho_\chi \!&=&\! b\,\Gamma_\phi\,\rho_\phi - \frac{\sv}{\epsilon_\chi} \left(\rho_\chi^2 - \rho_{\rm EQ}^2\right). \label{eq:boltzmann_chi}
\end{eqnarray}
Here, the equation of state relating the energy density and the pressure of the $\phi$ field is given by $\rho_\phi = p_\phi$, and translates into the term $3H(p_\phi + \rho_\phi) = 6H\rho_\phi$ in Eq.~\eqref{eq:boltzmann_phi}. The energy density and pressure in radiation are $\rho_R$ and $p_R = \rho_R/3$. At any time, the temperature $T$ is defined through the energy density in the relativistic component as $\rho_R = \alpha T^4$, with $\alpha = \pi^2g_*(T)/30$ and where $g_*(T)$ is the number of relativistic degrees of freedom. The WIMP number density $n_\chi$ is related to its energy density through $\rho_\chi = \epsilon_\chi n_\chi$, with the mean energy density per particle $\epsilon_\chi = \sqrt{m_\chi^2 + (c_1T)^2}$ and where we obtain $c_1 = 3.151$ through a numerical fit, confirming the results in Ref.~\cite{Redmond:2017tja}. The quantity $w = p_\chi/\rho_\chi \approx c_2T/\epsilon_\chi$, where $c_2=c_1/3$, tracks the change in the WIMP equation of state, so that $w = 0$ for $T \ll m_\chi$ and $w = 1/3$ for $T \gg m_\chi$. In the standard scenario, WIMPs reach thermal equilibrium and are relativistic as long as temperature is higher than $m_\chi$. Once temperature drops below their mass, the number density of WIMPs gets Boltzmann suppressed and eventually the number density in a comoving volume is fixed once the decoupling from the thermal bath occurs. This scenario might still hold when WIMPs are produced from the decay of the $\phi$ field, depending on the values of the parameters $b$, $\sv$, $m_\chi$. However, other scenarios in which WIMPs never reach thermal equilibrium are also possible. In the following, we assume that the number density of WIMPs is fixed prior $\TKIN$ and that WIMPs can be treated as non-relativistic, so $\TKIN \lesssim m_\chi$ for which we set $w = 0$ in Eq.~\eqref{eq:boltzmann_chi}. The energy density of WIMPs at chemical equilibrium is
\begin{equation}
	\rho_{\rm EQ} = \frac{g}{(2\pi)^3} \int E\,f(p)d^3p = \frac{g}{2\pi^2}\int_{m_\chi}^{+\infty} \frac{\sqrt{E^2-m_\chi^2}}{e^{E/T}+1} E^2dE,
\end{equation}
where $g$ is the number of WIMPs degrees of freedom. The set of Eqs.~\eqref{eq:boltzmann_phi}-\eqref{eq:boltzmann_chi} is closed when solved together with the Friedmann equation
\be
	H^2 = \frac{8\pi}{3\MP^2}\(\rho_\phi + \rho_R + \epsilon_\chi n_\chi\),
	\label{eq:friedmann}
\ee
where $\MP$ is the Planck mass. At early times $t \ll 1/\Gamma_\phi$, the Boltzmann Eq.~\eqref{eq:boltzmann_phi} and the Friedmann Eq.~\eqref{eq:friedmann} predict that, during KS, the energy density of the $\phi$ field and time scale as $\rho_\phi \sim a^{-6}$ and $t \sim a^3$, respectively. Contrarily to what found in kination models with negligible decay rate, for which $a \sim 1/T$~\cite{Spokoiny:1993kt, Joyce:1996cp, Salati:2002md, Profumo:2003hq}, in the model we study temperature depends on the scale factor as $T \propto \rho_\phi^{1/8} \propto a^{-3/4}$. In more details, even in the simplest case in which WIMPs are in chemical equilibrium, Eq.~\eqref{eq:boltzmann_R} can be reformulated as a differential equation describing the evolution of the entropy per comoving volume $s = (p_R + \rho_R)/T$, as
\begin{equation}
	\frac{ds}{dt} + 3Hs = (1-b)\frac{\Gamma_\phi}{T}\rho_\phi.
	\label{eq:entropy_evolution}
\end{equation}
As a consequence, entropy is not conserved in the model we consider because of the appearance of the dissipative term on the right hand side in Eq.~\eqref{eq:entropy_evolution}, coming from the decay of the $\phi$ field. We later confirm the dependence of temperature on the scale factor $T \propto a^{-3/4}$ by numerically solving the set of the Boltzmann equations, see Fig.~\ref{fig:plotbkgd} below. Notice that entropy conservation is not attained in Eq.~\eqref{eq:entropy_evolution} regardless of the details of the evolution of $\rho_\phi$.

We switch to the independent coordinates $x = m_\phi a$ and $\tau = \Gamma_\phi t$, while we write the dependent quantities in terms of the fields
\begin{equation}
	\Phi \!=\! Ax^6\rho_\phi / m_\phi,\quad R \!=\! Ax^4\rho_R/m_\phi,\quad X \!=\! A x^{3(1+w)}\rho_\chi/m_\phi.
	\label{eq:rescaledependent}
\end{equation}
Setting $Y = \Phi + x^2R + x^{3(1-w)}X$ and fixing the constant $A$ through the Friedmann equation,
\be
	\HH = \frac{1}{x}\frac{dx}{d\tau} =  \frac{H}{\Gamma_\phi} = \frac{\sqrt{Y}}{x^3},
\ee
we obtain that Eq.~\eqref{eq:friedmann} is recovered when
\be
	A = \frac{8\pi m_\phi}{3\MP^2\Gamma_\phi^2} \equiv \frac{m_\phi}{\rKIN},
\ee
where $\rKIN=\rho_R(\TKIN)$ is the energy density of the $\phi$ field at the temperature $\TKIN$ which, in the instantaneous thermalization approximation, is given by $\Gamma_\phi = H(\TKIN)$. With this definition, we rewrite the system of Eqs.~\eqref{eq:boltzmann_phi}-\eqref{eq:boltzmann_chi} as
\begin{eqnarray}
	\sqrt{Y}\,\Phi' &=& -x^2\Phi, \label{eq:boltzmann_phi1}\\
	\sqrt{Y}\,R' &=& (1-b)\Phi + s\left(X^2 - X_{\rm EQ}^2\right), \label{eq:boltzmann_R1} \\
	x\sqrt{Y}\,X' &=& b\Phi - s\left(X^2 - X_{\rm EQ}^2\right), \label{eq:boltzmann_chi1}
\end{eqnarray}
where $s = \rKIN\sv/\Gamma_\phi \epsilon_\chi$. This dimensionless form of the system can be easily extended to cosmologies other than KS. We assume that the relativistic species is always at equilibrium, so that temperature is related to $R$ as $T = \TKIN \,R^{1/4}/x$. At chemical equilibrium, the quantity $X_{\rm EQ} = x^{3}\rho_{\rm EQ}/\rKIN$ is
\bea
	X_{\rm EQ} &\approx& \frac{x^3}{\rKIN}\,m_\chi\,n_{\rm EQ},\\
	n_{\rm EQ} &=& g\(\frac{m_\chi\,T}{2\pi}\)^{3/2}\,e^{-m_\chi/T},
	\label{eq:X_equilibrium}
\eea
where the approximation has been obtained in the non-relativistic limit $T \ll m_\chi$. The set of Eqs.~\eqref{eq:boltzmann_phi1}-\eqref{eq:X_equilibrium} possesses a scaling symmetry,
\begin{equation}
	x \to \beta x, \quad \Phi \to \beta^6\Phi \quad R \to \beta^4R \quad X \to \beta^{3(1+w)}X,
\end{equation}
for any value of $\beta$, thanks to which the solution to the set of Boltzmann equations is independent on choice of the initial value $\Phi(x_I) = \Phi_I$ at $x = x_I$~\cite{Chung:1998rq, Giudice:2000ex}. We fix the initial condition by requiring that the Hubble rate at $x = x_I$ be $H_I = \sqrt{8\pi \rho_I/3\MP^2}$, where $\rho_I$ is the value of the energy density in the inflaton field at the inflation scale. Indeed, assuming that $\rho_R(x_I) \ll \rho_\phi(x_I) \equiv \rho_I$ gives
\begin{equation}
	\Phi(x) = \Phi_I \equiv \frac{\rho_I}{\rKIN}\,x_I^6,\quad R(x) = \sqrt{\Phi_I}\(x - x_I\).
\end{equation}
Thanks to this property, the actual value of the mass of the kination field $m_\phi$ does not have an impact on the solution of the equations, since $m_\phi$ enters the set of the Boltzmann equations only as an initial condition fixing the initial value of the Hubble rate. However, a value of $m_\phi \gtrsim m_\chi^2/T$ would be problematic for the cosmology at temperature $T$, since the relativistic byproducts would scatter off the thermal background and distort the relativistic spectra~\cite{Allahverdi:2002, Harigaya:2014waa}. The parameter space considered in this article is then $\TKIN < m_\chi \ll m_\phi \lesssim m_\chi^2/\TKIN$. Since the energy density $\rho_\phi$ during $\phi$-domination satisfies $\rho_\phi\sim a^{-6}$, the transition to the standard radiation-dominated scenario occurs when
\be
	\rho_{\phi,I}\,\left(\frac{x_I}{\xKIN}\right)^6 = \rKIN \equiv \alpha\,\TKIN^4.
	\label{eq:approximation}
\ee
Given the value of $\TKIN$, Eq.~\eqref{eq:approximation} defines the moment at which the $\phi$ field decays. For the illustrative purpose, we solve Eqs.~\eqref{eq:boltzmann_phi1} and~\eqref{eq:boltzmann_R1} for $\TKIN = 0.1\,$GeV, fixing the masses $m_\chi = 100\,$GeV and assuming that we can safely neglect the contribution of WIMPs to the total energy density. We have plot $\Phi$ (blue solid line) and $R$ (red dashed line) in units of $\Phi_I$, as well as the temperature $T/T_I$ (black dot-dashed line). The behavior of $\Phi$ confirms the scaling $\rho_\phi \sim a^{-6}$ and $\rho_R \sim a^{-3}$ at early times $x < \xKIN$, while for $x > \xKIN$ we obtain the scaling $\rho_R \sim a^{-4}$ for a radiation-dominated cosmology. The dashed vertical line marks the value of $\xKIN$ solution to Eq.~\eqref{eq:approximation}.
\begin{figure}[h!]
	\begin{center}
	\includegraphics[width=\linewidth]{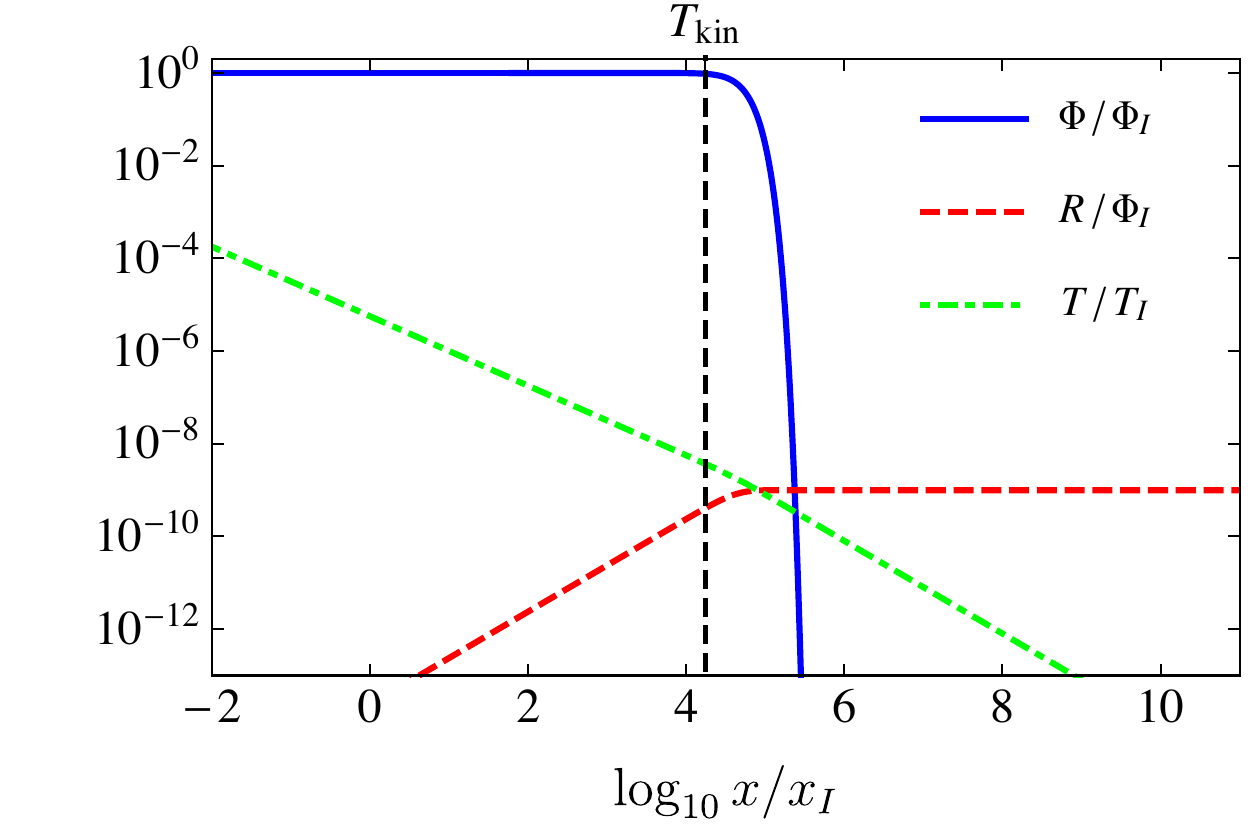}
	\caption{The quantities $\Phi$ and $R$, defined in Eq.~\eqref{eq:rescaledependent} and related to $\rho_\phi$ and $\rho_R$ respectively, in units of the initial value $\Phi_I$. The vertical dashed line marks the moment at which the transition to the standard scenario occurs, according to Eq.~\eqref{eq:approximation}. The green dot-dashed line shows the temperature $T(x)$, in units of its initial value $T(x_I)$.}
	\label{fig:plotbkgd}
	\end{center}
\end{figure}

\section{Production of WIMPs during kination}

With the framework presented in the previous Section, we solve the set of Boltzmann Eqs.~\eqref{eq:boltzmann_phi1}-\eqref{eq:boltzmann_chi1} for different values of $\TKIN$, to obtain the WIMP relic abundance $n_{\chi,{\rm fix}} \equiv n_\chi(T_{\rm fix})$ at some temperature $T_{\rm fix}$ at which the number density of WIMPs in a comoving volume is subsequently fixed by the cosmology. The present WIMP energy density in units of the critical density $\rho_c$ is then
\be
	\Omega_\chi = \frac{\rKIN}{\rho_c} \frac{g_S(T_0)}{g_S(T_{\rm fix})}\(\frac{T_0}{T_{\rm fix}}\)^3 \frac{X_{\rm fix}}{\sqrt{\Phi_I}},
	\label{eq:presentabundance}
\ee
where $X_{\rm fix} = x_{\rm fix}^3 m_\chi n_{\chi,{\rm fix}}/\rho_{\rm kin}$, $T_0$ is the present temperature of the radiation bath and $g_S(T)$ is the number of entropy degrees of freedom at temperature $T$.

\begin{figure}[h!]
	\begin{center}
	\includegraphics[width=\linewidth]{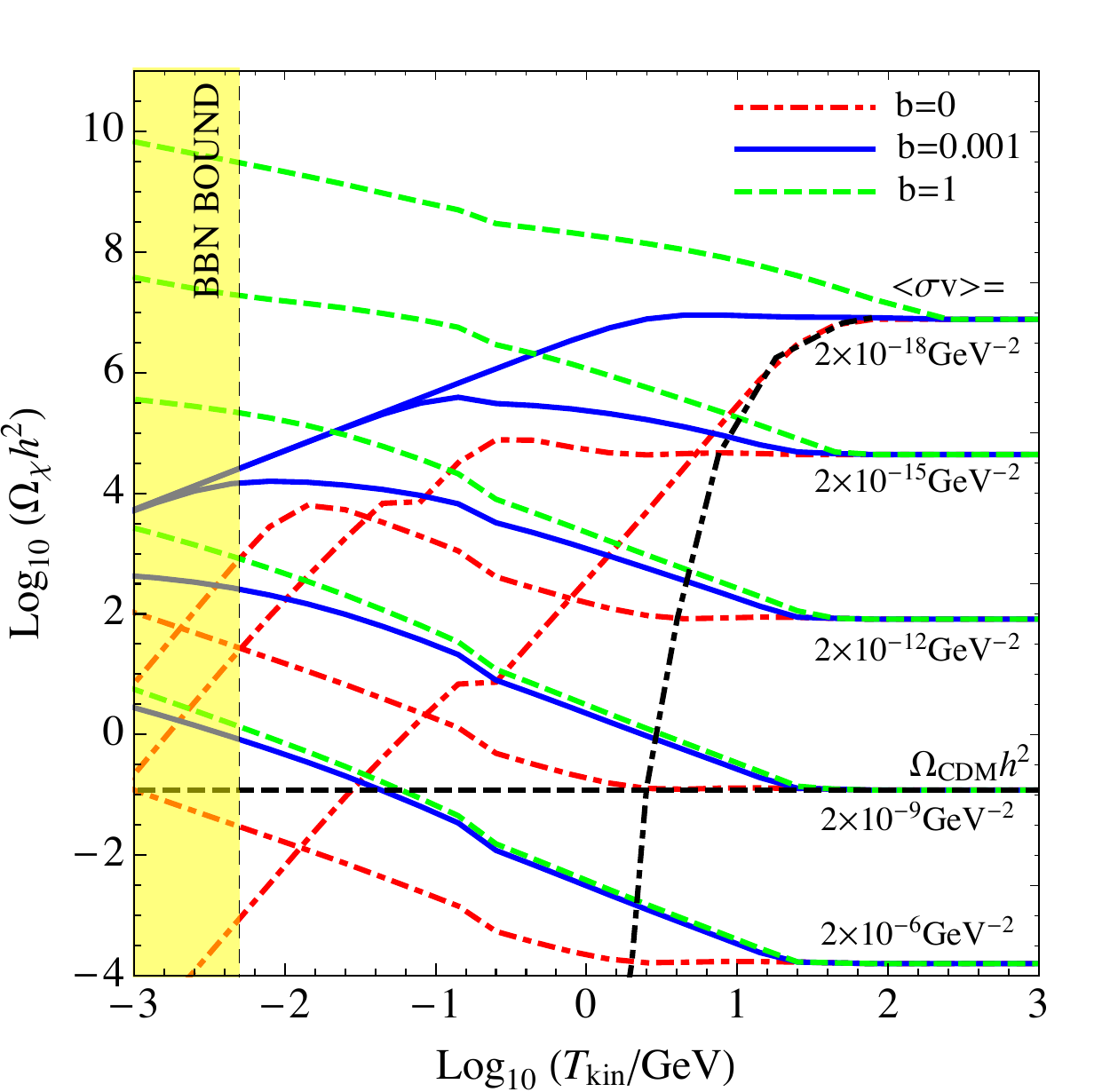}
	\caption{The present WIMP relic abundance for different values of $\sv$ (see figure labels), and for different values of the branching ratio: $b = 0$ (red dot-dashed line), $b = 0.001$ (blue solid line), and $b = 1$ (green dashed line), as a function of $\TKIN$. The horizontal dashed line shows the measured dark matter abundance $\Omega_{\rm DM}h^2 \sim 0.12$. The vertical yellow band defines the region excluded by BBN considerations, $\TKIN \geq 5\,$MeV. Freeze-out occurs in the kination scenario in the region to the left of the black dot-dashed curve.}
	\label{fig:omegavstrh}
	\end{center}
\end{figure}

We show different values of the abundance $\Omega_\chi h^2$ in Fig.~\ref{fig:omegavstrh}, as a function of the temperature $\TKIN$, the annihilation cross section $\sv$, and the branching ratio $b$. We have used three different values for the branching ratio: $b = 0$ (red dot-dashed line), $b = 0.001$ (blue solid line), and $b = 1$ (green dashed line), as well as five different values of $\sv$ ranging from $10^3\sv_{\rm std}$ to $10^{-9}\sv_{\rm std}$, where the value $\sv = \sv_{\rm std}$ gives the correct amount of WIMP dark matter from the freeze-out process in the standard cosmological scenario. The black dot-dashed curve marks the region where the freeze-out occurs in the kination scenario, lying to the left of the curve. We also show a portion of the region $\TKIN \gtrsim m_\chi$, where the freeze-out occurs in the standard scenario.

The numerical evaluation has been carried out by computing the value of $X$ at some later time $T_{\rm fix} \gg \TKIN$ in Eq.\eqref{eq:presentabundance} well into radiation-domination, using the conservation of entropy after $T_{\rm fix}$ to obtain the present abundance of WIMPs. To better understand the physics behind the Boltzmann equation for the evolution of $n_\chi$, we estimate the value of $X_{\rm fix}$ depending on the dominating mechanism of WIMP production. According to the values of $b$ and $\sv$, four possibilities appear, namely:
\begin{itemize}

\item {\bf Mechanism 1: Thermal production with chemical equilibrium (``freeze-out'').} We first focus on the case $b = 0$, corresponding to the negligible decay with respect to the annihilation into standard model particles and approximated by the red dot-dashed and blue solid lines in Fig.~\ref{fig:omegavstrh}. We assume that the temperature $\TKIN$ is lower than the freeze-out temperature $\Tfo$, which is defined as the temperature at which $n\sv = H$, or
\be
	n_{\rm EQ}(\Tfo) = \(\frac{\Tfo}{\TKIN}\)^4\frac{H(\TKIN)}{\sv}.
\ee
However, in modified cosmologies where the expansion rate is faster than during radiation (such as kination), WIMP annihilation persists even after the departure from chemical equilibrium (i.e. freeze-out) has occurred, actually ceasing when the Universe transitions to the standard radiation-dominated scenario~\cite{DEramo:2017gpl}. For this reason, the WIMP number density in the kination cosmology is not fixed at $\Tfo$ and annihilation continues until the temperature drops to $\TKIN$. We assume that the freeze-out is reached at $x_{\rm f.o.}$, when $X(x_{\rm f.o.}) = X_{\rm f.o.}$. At later times, using the approximation in Eq.~\eqref{eq:approximation} and the non-relativistic regime for the WIMPs, Eq.~\eqref{eq:boltzmann_chi1} reads
\be
	X' = -\frac{\rKIN\sv}{\Gamma_\phi m_\chi}\frac{X^2}{x\sqrt{\Phi_I}},
	\label{eq:boltzmann_chi_b0}
\ee
whose solution at $x > x_{\rm f.o.}$ give the abundance of thermally produced WIMPs at $\TKIN$ as
\bea
	X_{\rm kin, Th} &=& \[\frac{\sv \TKIN^2}{\sqrt{\Phi_I}} \frac{\MP}{m_\chi} \(\ln\frac{\xKIN}{x_{\rm fo}}+1\) + \frac{1}{X_{\rm f.o.}}\]^{-1},\quad \hbox{or},\\
	n_{\rm kin, Th} &=& \[\frac{\sv \TKIN^2\,\MP}{\rho_{\rm kin}} \(\ln\frac{\xKIN}{x_{\rm fo}}+1\) + \frac{x_{\rm kin}^3}{x_{\rm f.o.}^3}\frac{1}{n_{\chi, {\rm f.o.}}}\]^{-1}.
\eea
This expression has been obtained, for example, in Ref.~\cite{Redmond:2017tja}, since during kination $a_I^3H_I = a_{\rm kin}^3H\(T_{\rm kin}\)$. Neglecting $X_{\rm f.o.}$ and setting $T_{\rm fix} = \TKIN$ in Eq.~\eqref{eq:presentabundance}, the present abundance from the freeze-out mechanism gives
\be
	\Omega_{\chi, {\rm Th}} = \frac{g_S(T_0)}{g_S(\TKIN)}\frac{\rKIN}{\rho_c} \frac{m_\chi}{\sv} \frac{T_0^3}{\MP \TKIN^5} \(\ln\frac{\xKIN}{x_{\rm fo}} + 1\)^{-1} \propto \frac{1}{\TKIN}.
\ee
The solution describes, for example, the lines with negative slopes in Fig.~\ref{fig:omegavstrh} for $b=0$ and for the cross sections $\sv = 2\times 10^{-6}\,$GeV$^{-2}$, $\sv = 2\times 10^{-9}\,$GeV$^{-2}$, and $\sv = 2\times 10^{-12}\,$GeV$^{-2}$. In Fig.~\ref{fig:omegavstrh}, we have not approximated the numerical results by neglecting $X_{\rm f.o.}$.
\newline

\item {\bf Mechanism 2: Thermal production without ever reaching chemical equilibrium (``freeze-in'').} If the cross section is sufficiently low~\cite{Redmond:2017tja}, WIMPs never reach thermal equilibrium and their number density freezes in at a fixed quantity. Since the number density of particles is always smaller than their value at thermal equilibrium, we neglect $X \ll X_{\rm EQ}$ so Eq.~\eqref{eq:boltzmann_chi1} with $b=0$ reads
\be
	X' = \frac{\rKIN\sv}{\Gamma_\phi m_\chi}\frac{\left(X_{\rm EQ}^2\right)}{x\sqrt{\Phi_I}} = d_1 x^{11/4} \exp\(-2 d_2 x^{3/4}\),
	\label{eq:boltzmann_chi_fi}
\ee
where
\be
	d_1 \!=\! \frac{g^2\sv m_\chi^4 \TKIN^3}{(2\pi)^3\Gamma_\phi \rKIN \Phi_I^{1/8}}, \quad \hbox{and} \quad d_2 \!=\! \frac{m_\chi}{\TKIN \Phi_I^{1/8}}.
\ee
The solution to Eq.~\eqref{eq:boltzmann_chi_fi} reaches the asymptotic value of $X$ at freeze-in
\be
	X_{\rm kin, f.i.} = \frac{d_1}{d_2^5} = \frac{g^2\sv \TKIN^8\sqrt{\Phi_I}}{(2\pi)^3\Gamma_\phi \rKIN m_\chi},
\ee
which is reached when $x_{\rm fi} = (6/11d_2)^{4/3}$. This solution differs from the similar case of WIMP freeze-in studied in Ref.~\cite{Redmond:2017tja}, because of the different dependence of the temperature on the scale factor in this work. However, the methodology used is qualitatively the same. In principle, the number density in a comoving volume is fixed prior $\TKIN$, depending on the branching ratio of the $\phi$ field into radiation and the consequent entropy conservation during kination. However, we can safely set $T_{\rm fix} = \TKIN$ into Eq.~\eqref{eq:presentabundance} for this estimate, since the comoving number of WIMPs is conserved after freeze-in and $X_{\rm fix} = X_{\rm kin, f.i.}$. The present WIMP abundance from the freeze-in mechanism is then
\be
	\Omega_{\chi,{\rm f.i.}} = \frac{g_S(T_0)}{g_S(\TKIN)}\frac{g^2\sv T_0^3 \TKIN^5}{(2\pi)^3\Gamma_\phi \rho_c m_\chi} \propto \TKIN^3.
	\label{eq:presentabundance_freezein}
\ee
The solution describes, for example, the lines with positive slopes in Fig.~\ref{fig:omegavstrh}, for $b=0$ and the cross sections $\sv = 2\times 10^{-12}\,$GeV$^{-2}$, $\sv = 2\times 10^{-15}\,$GeV$^{-2}$, and $\sv = 2\times 10^{-18}\,$GeV$^{-2}$.
\newline

\item {\bf Mechanism 3: Non-thermal production without chemical equilibrium.} We now discuss the non-thermal production of dark matter, in the case in which the particle has never reached the chemical equilibrium. For a sufficiently large branching ratio $b$ and for $\TKIN \ll m_\chi$, the abundance of dark matter is set by the decay of the $\phi$ field, with an energy density at $\TKIN$ given by~\cite{Gelmini:2006pw, Choi:2008zq, Acharya:2009zt, Baer:2014eja, Kane:2015qea, Kim:2016spf}
\be
	\rho_\chi(\TKIN) \approx b\,\rho_\phi(\TKIN).
	\label{eq:nonthermalnoeq}
\ee
Deriving the result from directly integrating Eq.~\eqref{eq:boltzmann_chi1} with $\sv = 0$ and neglecting the contributions from $R$ and $X$ in the denominator gives an extra logarithmic dependence on $\xKIN$, as
\be
	X_{\rm kin, decay} = b\sqrt{\Phi_I} \ln\frac{\xKIN}{x_I}.
	\label{eq:presentdensitynonthermal}
\ee
The present WIMP abundance when the non-thermal production dominates is given by Eq.~\eqref{eq:presentabundance} with $T_{\rm fix} = \TKIN$, corresponding to the moment at which WIMPs are produced from the decay of the $\phi$ field with the initial amount in Eq.~\eqref{eq:nonthermalnoeq},
\be
	\Omega_{\chi,{\rm decay}} = \frac{b\rKIN}{\rho_c} \frac{g_S(T_0)}{g_S(\TKIN)}\(\frac{T_0}{\TKIN}\)^3 \ln\frac{\xKIN}{x_I}.
	\label{eq:presentabundance_m3}
\ee
This latter expression predicts the behavior $\Omega_{\chi,{\rm decay}} \propto \TKIN$ corresponding to $b = 1$ (green dashed line) or $b = 0.001$ (blue solid line) with $\sv = 10^{-18}\,$GeV$^{-2}$ in Fig.~\ref{fig:omegavstrh}, for $\TKIN \lesssim 10\,$GeV. Notice that, except for the logarithmic dependence which is present in the kination cosmology, the result in Eq.~\eqref{eq:presentdensitynonthermal} is independent of the cosmology used.
\newline

\item {\bf Mechanism 4: Non-thermal production with chemical equilibrium.} If the branching ratio $b$ is sufficiently high, the evolution of the WIMP number density attains a secular equilibrium in which the rate at which WIMPs are produced from the decay of the $\phi$ field equates that from WIMP annihilation. In this regime, the quantity $X$ is fixed to the value obtained by setting to zero the right-hand side of Eq.~\eqref{eq:boltzmann_chi1},
\be
	X_{\rm kin, sec}  = \sqrt{\frac{b\Phi_I \Gamma_\phi m_\chi}{\rKIN\sv}}.
	\label{eq:Xkin}
\ee
The result in Eq.~\eqref{eq:Xkin}, confirmed numerically in Fig.~\ref{fig:plotrecap} below, can be alternatively derived by considering the balancing between the decay rate of the $\phi$ field into WIMPs and the annihilation rate of WIMPs, valid at $\TKIN$ when $\rho_\phi = \rho_{\rm kin}$, as
\be
	\(m_\chi\,n_\chi\)\( \sv n_\chi\) = b\Gamma_\phi \rho_\phi.
	\label{eq:def_xb}
\ee
The value of $X_{\rm kin, sec}$ remains constant until $\TKIN$, without experiencing the additional depletion obtained in the freeze-out regime with a faster-than-radiation expansion rate~\cite{DEramo:2017gpl, Redmond:2017tja}. However, when the temperature of the plasma falls below $\TKIN$, the secular equilibrium is no longer maintained since the energy density in the $\phi$ field drops to zero and WIMPs are no longer produced. In this new regime, radiation evolves as $R(x) = \Phi_I\(x/\xKIN\)^2$ and Eq.~\eqref{eq:boltzmann_chi1} reads
\be
	\frac{dX}{d(x/\xKIN)} = -\frac{\rKIN\sv}{\Gamma_\phi m_\chi} \frac{X^2}{\(\frac{x}{\xKIN}\)^2\sqrt{\Phi_I}}, \label{eq:boltzmann_chi2}
\ee
which gives the WIMP number density $X = \sqrt{\Phi_I}\Gamma_\phi m_\chi/\sv\rho_{\xKIN}$ when $x \gg \xKIN$, corresponding to the number density $n_\chi = \Gamma_\phi/\sv$. Compare Eq.~\eqref{eq:boltzmann_chi2} with the approximation in Eq.~\eqref{eq:boltzmann_chi_fi} valid during $\phi$-domination and $b = 0$, the expression for Eq.~\eqref{eq:boltzmann_chi2} has been obtained by replacing $\rho_\phi$ with $\rho_R$ as the leading term in the expression for the Hubble rate. The present abundance is expressed as
\be
	\Omega_{\chi,{\rm nonTh}} = \frac{m_\chi \Gamma_\phi}{\rho_c\,\sv} \frac{g_S(T_0)}{g_S(\TKIN)} \(\frac{T_0}{\TKIN}\)^3 \propto \frac{1}{\TKIN}.
	\label{eq:presentdensitynonthermal1}
\ee
\end{itemize}

\section{Discussion and summary}

\begin{figure}[h!]
	\begin{center}
	\includegraphics[width=\linewidth]{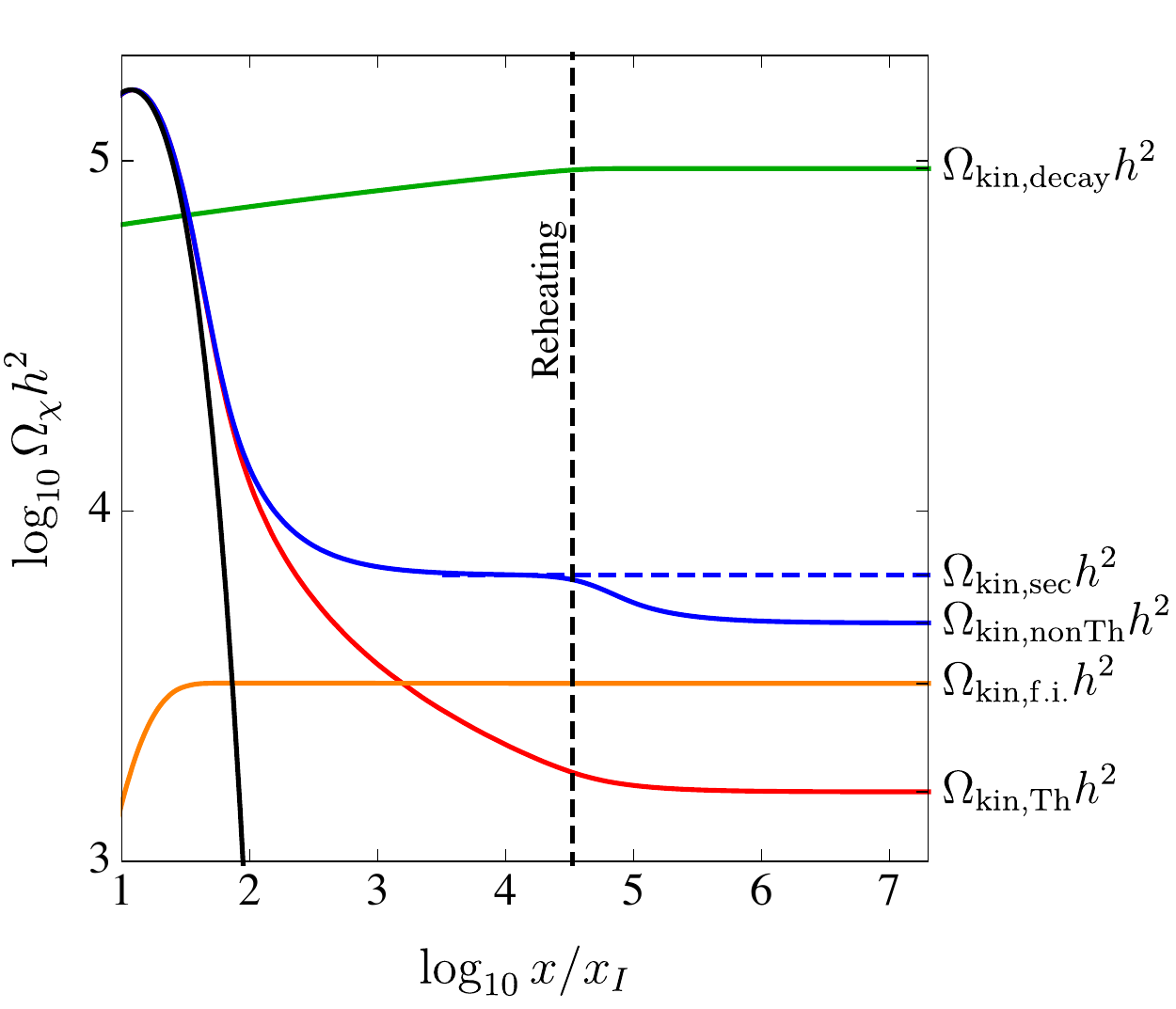}
	\caption{The solution to the Boltzmann Eq.~\eqref{eq:boltzmann_chi1} for different values of the annihilation cross section. We have set $\TKIN=100\,$MeV and, when non-thermal production is considered, $b=2\times 10^{-4}$. See text for further details.}
	\label{fig:plotrecap}
	\end{center}
\end{figure}

In Fig.~\ref{fig:plotrecap}, we summarize the results obtained by considering the behavior of the quantity $X$ solution to the Boltzmann Eq.~\eqref{eq:boltzmann_chi1} for $\TKIN = 100\,$MeV, $m_\chi = 100\,$GeV, and for different values of $b$ and $\sv$ so that each of the mechanisms discussed above is attained. In more detail, we consider $b = 0, \sv = 2 \times 10^{-12}{\rm \,GeV^{-2}}$ (describing thermal production with chemical equilibrium or Mechanism 1, red line), $b = 0, \sv = 2 \times 10^{-18}{\rm \,GeV^{-2}}$ (describing thermal production without chemical equilibrium or Mechanism 2, orange line), $b = 2\times 10^{-4}, \sv = 2 \times 10^{-18}{\rm \,GeV^{-2}}$ (describing non-thermal production without chemical equilibrium or Mechanism 3, green line), and $b = 2\times 10^{-4}, \sv = 2 \times 10^{-9}{\rm \,GeV^{-2}}$ (describing non-thermal production with chemical equilibrium or Mechanism 4, blue line). The black solid line in Fig.~\ref{fig:plotrecap} represents $X_{\rm EQ}$ for $\TKIN = 0.1\,$GeV.
In addition to the thermal mechanisms of production recently discussed in Refs.~\cite{DEramo:2017gpl, Redmond:2017tja}, we have included the possibility that the field $\phi$ responsible for the kination period decays into radiation and WIMPs. We have then studied the non-thermal production of WIMPs in the kination cosmology, as summarized in Fig.~\ref{fig:omegavstrh}. Given the value $\sv_{\rm std} = 2\times 10^{-9}\,$GeV$^{-2}$ that gives the present abundance of DM from the freeze-out of WIMPs in the standard cosmology, Fig.~\ref{fig:omegavstrh} shows that larger values of $\sv$ can still lead to the right DM abundance, if WIMPs are produced either through Mechanisms 1 or Mechanism 4 during the KS. Similarly, we can have $\sv$ smaller than its standard value and still have the correct amount of DM, if WIMPs are produced through Mechanism 2. Non-thermal production without chemical equilibrium (Mechanism 3) would lead to the correct amount of DM only for values of $\TKIN$ that are excluded by the BBN considerations. Using the bound $\TKIN \gtrsim 5\,$MeV, we infer the possible range of the parameter
\be
	4 \times 10^{-16}{\rm \,GeV^{-2}} \lesssim \sv \lesssim 2 \times 10^{-5} {\rm \,GeV^{-2}},
	\label{eq:bound_sigma}
\ee
the lower bound being obtained by using Mechanism 2 and the upper bound being given by Mechanism 4 with $b = 1$. This result is valid for a WIMP of mass $m_\chi = 100\,$GeV. If WIMPs annihilate predominantly into a pair of photons, the annihilation cross section is bound by the recent measurements by the FERMI-Large Area Telescope satellite~\cite{Ackermann:2015lka} as $\sv \lesssim 10^{-10}{\rm \,GeV}^{-2}$. However, notice that the cross section constrained in Eq.~\eqref{eq:bound_sigma} refer to different epochs with respect to the constraints by the FERMI satellite.

\begin{figure}[tb]
	\begin{center}
	\includegraphics[width=\linewidth]{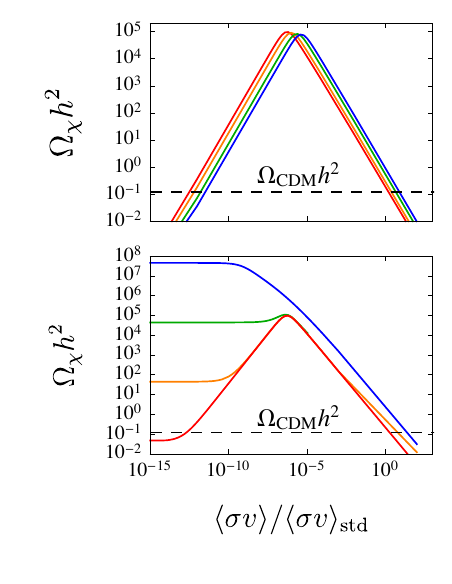}
	\caption{Top panel: the present relic density $\Omega_\chi h^2$, in units of the critical density of the universe, as a function of the quantity $\sv$ for $\TKIN = 1\,$GeV, $b = 0$, and for the value of the WIMP mass $m_\chi = 100\,$GeV (red line), $m_\chi = 200\,$GeV (orange line), $m_\chi = 500\,$GeV (green line), and $m_\chi = 1000\,$GeV (blue line). Bottom panel: same as the top panel for $\TKIN = 1\,$GeV, $m_\chi = 100\,$GeV and for the branching ratio $b = 10^{-11}$ (red line),  $b = 10^{-8}$ (orange line),  $b = 10^{-5}$ (green line), and $b = 10^{-2}$ (blue line).}
	\label{fig:plot4}
	\end{center}
\end{figure}

The dependence of the solution on the parameters $b$, $\sv$, and $m_\chi$ is also depicted in Fig.~\ref{fig:plot4}, where the top panel shows how the relic density $\Omega_\chi h^2$ depends on the value of $\sv$ for $b = 0$, $\TKIN = 1\,$GeV and for different values of the WIMP mass: $m_\chi = 100\,$GeV (red line), $m_\chi = 200\,$GeV (orange line), $m_\chi = 500\,$GeV (green line), and $m_\chi = 1000\,$GeV (blue line). This result, corresponding to the usual thermal production during kination~\cite{Redmond:2017tja}, shows that the correct relic density can be attained through either the freeze-out mechanism, for $\sv \sim 10^{-10}{\rm \,GeV^{-2}}$, or the freeze-in mechanism, for $\sv \sim 10^{-21}{\rm \,GeV^{-2}}$, the exact values depending on the values of $m_\chi$ and $\TKIN$. The new result is presented in the bottom panel of Fig.~\ref{fig:plot4}, which shows the dependence of the present relic density on $\sv$, for $m_\chi = 100\,$GeV and for different values of the branching ratio. We set $b = 10^{-11}$ (red line),  $b = 10^{-8}$ (orange line),  $b = 10^{-5}$ (green line), and $b = 10^{-2}$ (blue line). For a fixed value of $b$, the relic density reaches a constant value when the the quantity $\sv$ is decreased, since for a sufficiently low value of $\sv$ the production of WIMPs from the decay of the $\phi$ field overcomes the depletion within the freeze-in mechanism. In other words, when $\Omega_\chi h^2$ has reached a plateau, the number of WIMPs has been fixed by the details of Mechanism 4. For $b > \bar{b}$, the value of this plateau is too large to explain the observed amount of dark matter, with a consequent overclosure of the universe. For the parameters chosen, we find $\bar{b} = 3\times 10^{-10}$. In this scenario, only values of the order of $\sv = \mathcal{O}\(1-10\)\sv_{\rm std}$ yield the correct value of $\Omega_\chi h^2$.

To summarize, if the DM is a WIMP of mass $m_\chi = 100\,$GeV, and if the annihilation cross section is measured to lie outside of the bound in Eq.~\eqref{eq:bound_sigma}, then the kination model discussed would have to be discarded. The same analysis can be performed by varying the masses of the WIMP and of the $\phi$ field, which would lead to different values of the bounds in Eq.~\eqref{eq:bound_sigma}. The results presented have been obtained assuming a kination scenario with equation of state $p_\phi = \rho\phi$; anyhow, any faster-than-radiation expansion rate for the background, described by a fluid with with equation of state $p_\phi/\rho_\phi > 1/3$ would yield to results that are qualitatively similar, as shown in Ref.~\cite{DEramo:2017gpl} for the case $\Gamma_\phi = 0$. If ever discovered, the properties of the WIMP could then shed light on the pre-BBN cosmology.

\acknowledgments{We thank Adrienne Erickcek, Kayla Redmond, and Sunny Vagnozzi for reading earlier versions of the manuscript. We acknowledge support by the Vetenskapsr\r{a}det (Swedish Research Council) through contract No. 638-2013-8993 and the Oskar Klein Centre for Cosmoparticle Physics.}

\bibliography{NonThermalWIMPs}

\begin{thebibliography}{82}%
\makeatletter
\providecommand \@ifxundefined [1]{%
 \@ifx{#1\undefined}
}%
\providecommand \@ifnum [1]{%
 \ifnum #1\expandafter \@firstoftwo
 \else \expandafter \@secondoftwo
 \fi
}%
\providecommand \@ifx [1]{%
 \ifx #1\expandafter \@firstoftwo
 \else \expandafter \@secondoftwo
 \fi
}%
\providecommand \natexlab [1]{#1}%
\providecommand \enquote  [1]{``#1''}%
\providecommand \bibnamefont  [1]{#1}%
\providecommand \bibfnamefont [1]{#1}%
\providecommand \citenamefont [1]{#1}%
\providecommand \href@noop [0]{\@secondoftwo}%
\providecommand \href [0]{\begingroup \@sanitize@url \@href}%
\providecommand \@href[1]{\@@startlink{#1}\@@href}%
\providecommand \@@href[1]{\endgroup#1\@@endlink}%
\providecommand \@sanitize@url [0]{\catcode `\\12\catcode `\$12\catcode
  `\&12\catcode `\#12\catcode `\^12\catcode `\_12\catcode `\%12\relax}%
\providecommand \@@startlink[1]{}%
\providecommand \@@endlink[0]{}%
\providecommand \url  [0]{\begingroup\@sanitize@url \@url }%
\providecommand \@url [1]{\endgroup\@href {#1}{\urlprefix }}%
\providecommand \urlprefix  [0]{URL }%
\providecommand \Eprint [0]{\href }%
\providecommand \doibase [0]{http://dx.doi.org/}%
\providecommand \selectlanguage [0]{\@gobble}%
\providecommand \bibinfo  [0]{\@secondoftwo}%
\providecommand \bibfield  [0]{\@secondoftwo}%
\providecommand \translation [1]{[#1]}%
\providecommand \BibitemOpen [0]{}%
\providecommand \bibitemStop [0]{}%
\providecommand \bibitemNoStop [0]{.\EOS\space}%
\providecommand \EOS [0]{\spacefactor3000\relax}%
\providecommand \BibitemShut  [1]{\csname bibitem#1\endcsname}%
\let\auto@bib@innerbib\@empty
\bibitem [{\citenamefont {Jungman}\ \emph {et~al.}(1996)\citenamefont
  {Jungman}, \citenamefont {Kamionkowski},\ and\ \citenamefont
  {Griest}}]{Jungman:1995df}%
  \BibitemOpen
  \bibfield  {author} {\bibinfo {author} {\bibfnamefont {G.}~\bibnamefont
  {Jungman}}, \bibinfo {author} {\bibfnamefont {M.}~\bibnamefont
  {Kamionkowski}}, \ and\ \bibinfo {author} {\bibfnamefont {K.}~\bibnamefont
  {Griest}},\ }\href {\doibase 10.1016/0370-1573(95)00058-5} {\bibfield
  {journal} {\bibinfo  {journal} {Phys. Rept.}\ }\textbf {\bibinfo {volume}
  {267}},\ \bibinfo {pages} {195} (\bibinfo {year} {1996})},\ \Eprint
  {http://arxiv.org/abs/hep-ph/9506380} {arXiv:hep-ph/9506380 [hep-ph]}
  \BibitemShut {NoStop}%
\bibitem [{\citenamefont {Bertone}\ \emph {et~al.}(2005)\citenamefont
  {Bertone}, \citenamefont {Hooper},\ and\ \citenamefont
  {Silk}}]{Bertone:2004pz}%
  \BibitemOpen
  \bibfield  {author} {\bibinfo {author} {\bibfnamefont {G.}~\bibnamefont
  {Bertone}}, \bibinfo {author} {\bibfnamefont {D.}~\bibnamefont {Hooper}}, \
  and\ \bibinfo {author} {\bibfnamefont {J.}~\bibnamefont {Silk}},\ }\href
  {\doibase 10.1016/j.physrep.2004.08.031} {\bibfield  {journal} {\bibinfo
  {journal} {Phys. Rept.}\ }\textbf {\bibinfo {volume} {405}},\ \bibinfo
  {pages} {279} (\bibinfo {year} {2005})},\ \Eprint
  {http://arxiv.org/abs/hep-ph/0404175} {arXiv:hep-ph/0404175 [hep-ph]}
  \BibitemShut {NoStop}%
\bibitem [{\citenamefont {Ade}\ \emph {et~al.}(2015)\citenamefont {Ade} \emph
  {et~al.}}]{Ade:2015tva}%
  \BibitemOpen
  \bibfield  {author} {\bibinfo {author} {\bibfnamefont {P.~A.~R.}\
  \bibnamefont {Ade}} \emph {et~al.} (\bibinfo {collaboration} {BICEP2,
  Planck}),\ }\href {\doibase 10.1103/PhysRevLett.114.101301} {\bibfield
  {journal} {\bibinfo  {journal} {Phys. Rev. Lett.}\ }\textbf {\bibinfo
  {volume} {114}},\ \bibinfo {pages} {101301} (\bibinfo {year} {2015})},\
  \Eprint {http://arxiv.org/abs/1502.00612} {arXiv:1502.00612 [astro-ph.CO]}
  \BibitemShut {NoStop}%
\bibitem [{\citenamefont {Ade}\ \emph {et~al.}(2016)\citenamefont {Ade} \emph
  {et~al.}}]{Ade:2015xua}%
  \BibitemOpen
  \bibfield  {author} {\bibinfo {author} {\bibfnamefont {P.~A.~R.}\
  \bibnamefont {Ade}} \emph {et~al.} (\bibinfo {collaboration} {Planck}),\
  }\href {\doibase 10.1051/0004-6361/201525830} {\bibfield  {journal} {\bibinfo
   {journal} {Astron. Astrophys.}\ }\textbf {\bibinfo {volume} {594}},\
  \bibinfo {pages} {A13} (\bibinfo {year} {2016})},\ \Eprint
  {http://arxiv.org/abs/1502.01589} {arXiv:1502.01589 [astro-ph.CO]}
  \BibitemShut {NoStop}%
\bibitem [{\citenamefont {Vysotsky}\ \emph {et~al.}(1977)\citenamefont
  {Vysotsky}, \citenamefont {Dolgov},\ and\ \citenamefont
  {Zeldovich}}]{Vysotsky:1977pe}%
  \BibitemOpen
  \bibfield  {author} {\bibinfo {author} {\bibfnamefont {M.~I.}\ \bibnamefont
  {Vysotsky}}, \bibinfo {author} {\bibfnamefont {A.~D.}\ \bibnamefont
  {Dolgov}}, \ and\ \bibinfo {author} {\bibfnamefont {{\relax Ya}.~B.}\
  \bibnamefont {Zeldovich}},\ }\href@noop {} {\bibfield  {journal} {\bibinfo
  {journal} {JETP Lett.}\ }\textbf {\bibinfo {volume} {26}},\ \bibinfo {pages}
  {188} (\bibinfo {year} {1977})},\ \bibinfo {note} {[Pisma Zh. Eksp. Teor.
  Fiz.26,200(1977)]}\BibitemShut {NoStop}%
\bibitem [{\citenamefont {Hut}(1977)}]{Hut:1977zn}%
  \BibitemOpen
  \bibfield  {author} {\bibinfo {author} {\bibfnamefont {P.}~\bibnamefont
  {Hut}},\ }\href {\doibase 10.1016/0370-2693(77)90139-3} {\bibfield  {journal}
  {\bibinfo  {journal} {Phys. Lett.}\ }\textbf {\bibinfo {volume} {69B}},\
  \bibinfo {pages} {85} (\bibinfo {year} {1977})}\BibitemShut {NoStop}%
\bibitem [{\citenamefont {Sato}\ and\ \citenamefont
  {Kobayashi}(1977)}]{Sato:1977ye}%
  \BibitemOpen
  \bibfield  {author} {\bibinfo {author} {\bibfnamefont {K.}~\bibnamefont
  {Sato}}\ and\ \bibinfo {author} {\bibfnamefont {M.}~\bibnamefont
  {Kobayashi}},\ }\bibfield  {booktitle} {\emph {\bibinfo {booktitle} {{IN
  *TOKYO 1977, PROCEEDINGS, NEW PARTICLES, THE STRUCTURE OF HADRONS, AND WEAK
  NEUTRAL CURRENTS* 163-167.}}},\ }\href {\doibase 10.1143/PTP.58.1775}
  {\bibfield  {journal} {\bibinfo  {journal} {Prog. Theor. Phys.}\ }\textbf
  {\bibinfo {volume} {58}},\ \bibinfo {pages} {1775} (\bibinfo {year}
  {1977})}\BibitemShut {NoStop}%
\bibitem [{\citenamefont {Lee}\ and\ \citenamefont
  {Weinberg}(1977)}]{Lee:1977ua}%
  \BibitemOpen
  \bibfield  {author} {\bibinfo {author} {\bibfnamefont {B.~W.}\ \bibnamefont
  {Lee}}\ and\ \bibinfo {author} {\bibfnamefont {S.}~\bibnamefont {Weinberg}},\
  }\href {\doibase 10.1103/PhysRevLett.39.165} {\bibfield  {journal} {\bibinfo
  {journal} {Phys. Rev. Lett.}\ }\textbf {\bibinfo {volume} {39}},\ \bibinfo
  {pages} {165} (\bibinfo {year} {1977})}\BibitemShut {NoStop}%
\bibitem [{\citenamefont {Dicus}\ \emph {et~al.}(1978)\citenamefont {Dicus},
  \citenamefont {Kolb},\ and\ \citenamefont {Teplitz}}]{Dicus:1977qy}%
  \BibitemOpen
  \bibfield  {author} {\bibinfo {author} {\bibfnamefont {D.~A.}\ \bibnamefont
  {Dicus}}, \bibinfo {author} {\bibfnamefont {E.~W.}\ \bibnamefont {Kolb}}, \
  and\ \bibinfo {author} {\bibfnamefont {V.~L.}\ \bibnamefont {Teplitz}},\
  }\href {\doibase 10.1086/156031} {\bibfield  {journal} {\bibinfo  {journal}
  {Astrophys. J.}\ }\textbf {\bibinfo {volume} {221}},\ \bibinfo {pages} {327}
  (\bibinfo {year} {1978})}\BibitemShut {NoStop}%
\bibitem [{\citenamefont {Steigman}(1979)}]{Steigman:1979kw}%
  \BibitemOpen
  \bibfield  {author} {\bibinfo {author} {\bibfnamefont {G.}~\bibnamefont
  {Steigman}},\ }\href {\doibase 10.1146/annurev.ns.29.120179.001525}
  {\bibfield  {journal} {\bibinfo  {journal} {Ann. Rev. Nucl. Part. Sci.}\
  }\textbf {\bibinfo {volume} {29}},\ \bibinfo {pages} {313} (\bibinfo {year}
  {1979})}\BibitemShut {NoStop}%
\bibitem [{\citenamefont {Bernstein}\ \emph {et~al.}(1985)\citenamefont
  {Bernstein}, \citenamefont {Brown},\ and\ \citenamefont
  {Feinberg}}]{Bernstein:1985th}%
  \BibitemOpen
  \bibfield  {author} {\bibinfo {author} {\bibfnamefont {J.}~\bibnamefont
  {Bernstein}}, \bibinfo {author} {\bibfnamefont {L.~S.}\ \bibnamefont
  {Brown}}, \ and\ \bibinfo {author} {\bibfnamefont {G.}~\bibnamefont
  {Feinberg}},\ }\href {\doibase 10.1103/PhysRevD.32.3261} {\bibfield
  {journal} {\bibinfo  {journal} {Phys. Rev.}\ }\textbf {\bibinfo {volume}
  {D32}},\ \bibinfo {pages} {3261} (\bibinfo {year} {1985})}\BibitemShut
  {NoStop}%
\bibitem [{\citenamefont {Kolb}\ and\ \citenamefont
  {Olive}(1986)}]{Kolb:1985nn}%
  \BibitemOpen
  \bibfield  {author} {\bibinfo {author} {\bibfnamefont {E.~W.}\ \bibnamefont
  {Kolb}}\ and\ \bibinfo {author} {\bibfnamefont {K.~A.}\ \bibnamefont
  {Olive}},\ }\href {\doibase 10.1103/PhysRevD.33.1202,
  10.1103/PhysRevD.34.2531} {\bibfield  {journal} {\bibinfo  {journal} {Phys.
  Rev.}\ }\textbf {\bibinfo {volume} {D33}},\ \bibinfo {pages} {1202} (\bibinfo
  {year} {1986})},\ \bibinfo {note} {[Erratum: Phys.
  Rev.D34,2531(1986)]}\BibitemShut {NoStop}%
\bibitem [{\citenamefont {Profumo}\ \emph {et~al.}(2006)\citenamefont
  {Profumo}, \citenamefont {Sigurdson},\ and\ \citenamefont
  {Kamionkowski}}]{Profumo:2006bv}%
  \BibitemOpen
  \bibfield  {author} {\bibinfo {author} {\bibfnamefont {S.}~\bibnamefont
  {Profumo}}, \bibinfo {author} {\bibfnamefont {K.}~\bibnamefont {Sigurdson}},
  \ and\ \bibinfo {author} {\bibfnamefont {M.}~\bibnamefont {Kamionkowski}},\
  }\href {\doibase 10.1103/PhysRevLett.97.031301} {\bibfield  {journal}
  {\bibinfo  {journal} {Phys. Rev. Lett.}\ }\textbf {\bibinfo {volume} {97}},\
  \bibinfo {pages} {031301} (\bibinfo {year} {2006})},\ \Eprint
  {http://arxiv.org/abs/astro-ph/0603373} {arXiv:astro-ph/0603373 [astro-ph]}
  \BibitemShut {NoStop}%
\bibitem [{\citenamefont {Griest}\ and\ \citenamefont
  {Seckel}(1991)}]{Griest:1990kh}%
  \BibitemOpen
  \bibfield  {author} {\bibinfo {author} {\bibfnamefont {K.}~\bibnamefont
  {Griest}}\ and\ \bibinfo {author} {\bibfnamefont {D.}~\bibnamefont
  {Seckel}},\ }\href {\doibase 10.1103/PhysRevD.43.3191} {\bibfield  {journal}
  {\bibinfo  {journal} {Phys. Rev.}\ }\textbf {\bibinfo {volume} {D43}},\
  \bibinfo {pages} {3191} (\bibinfo {year} {1991})}\BibitemShut {NoStop}%
\bibitem [{\citenamefont {Edsjo}\ and\ \citenamefont
  {Gondolo}(1997)}]{Edsjo:1997bg}%
  \BibitemOpen
  \bibfield  {author} {\bibinfo {author} {\bibfnamefont {J.}~\bibnamefont
  {Edsjo}}\ and\ \bibinfo {author} {\bibfnamefont {P.}~\bibnamefont
  {Gondolo}},\ }\href {\doibase 10.1103/PhysRevD.56.1879} {\bibfield  {journal}
  {\bibinfo  {journal} {Phys. Rev.}\ }\textbf {\bibinfo {volume} {D56}},\
  \bibinfo {pages} {1879} (\bibinfo {year} {1997})},\ \Eprint
  {http://arxiv.org/abs/hep-ph/9704361} {arXiv:hep-ph/9704361 [hep-ph]}
  \BibitemShut {NoStop}%
\bibitem [{\citenamefont {D'Agnolo}\ and\ \citenamefont
  {Ruderman}(2015)}]{DAgnolo:2015ujb}%
  \BibitemOpen
  \bibfield  {author} {\bibinfo {author} {\bibfnamefont {R.~T.}\ \bibnamefont
  {D'Agnolo}}\ and\ \bibinfo {author} {\bibfnamefont {J.~T.}\ \bibnamefont
  {Ruderman}},\ }\href {\doibase 10.1103/PhysRevLett.115.061301} {\bibfield
  {journal} {\bibinfo  {journal} {Phys. Rev. Lett.}\ }\textbf {\bibinfo
  {volume} {115}},\ \bibinfo {pages} {061301} (\bibinfo {year} {2015})},\
  \Eprint {http://arxiv.org/abs/1505.07107} {arXiv:1505.07107 [hep-ph]}
  \BibitemShut {NoStop}%
\bibitem [{\citenamefont {Cline}\ \emph {et~al.}(2017)\citenamefont {Cline},
  \citenamefont {Liu}, \citenamefont {Slatyer},\ and\ \citenamefont
  {Xue}}]{Cline:2017tka}%
  \BibitemOpen
  \bibfield  {author} {\bibinfo {author} {\bibfnamefont {J.}~\bibnamefont
  {Cline}}, \bibinfo {author} {\bibfnamefont {H.}~\bibnamefont {Liu}}, \bibinfo
  {author} {\bibfnamefont {T.}~\bibnamefont {Slatyer}}, \ and\ \bibinfo
  {author} {\bibfnamefont {W.}~\bibnamefont {Xue}},\ }\href {\doibase
  10.1103/PhysRevD.96.083521} {\bibfield  {journal} {\bibinfo  {journal} {Phys.
  Rev.}\ }\textbf {\bibinfo {volume} {D96}},\ \bibinfo {pages} {083521}
  (\bibinfo {year} {2017})},\ \Eprint {http://arxiv.org/abs/1702.07716}
  {arXiv:1702.07716 [hep-ph]} \BibitemShut {NoStop}%
\bibitem [{\citenamefont {Chang}\ \emph {et~al.}(2010)\citenamefont {Chang},
  \citenamefont {Pierce},\ and\ \citenamefont {Weiner}}]{Chang:2009yt}%
  \BibitemOpen
  \bibfield  {author} {\bibinfo {author} {\bibfnamefont {S.}~\bibnamefont
  {Chang}}, \bibinfo {author} {\bibfnamefont {A.}~\bibnamefont {Pierce}}, \
  and\ \bibinfo {author} {\bibfnamefont {N.}~\bibnamefont {Weiner}},\ }\href
  {\doibase 10.1088/1475-7516/2010/01/006} {\bibfield  {journal} {\bibinfo
  {journal} {JCAP}\ }\textbf {\bibinfo {volume} {1001}},\ \bibinfo {pages}
  {006} (\bibinfo {year} {2010})},\ \Eprint {http://arxiv.org/abs/0908.3192}
  {arXiv:0908.3192 [hep-ph]} \BibitemShut {NoStop}%
\bibitem [{\citenamefont {Fan}\ \emph {et~al.}(2010)\citenamefont {Fan},
  \citenamefont {Reece},\ and\ \citenamefont {Wang}}]{Fan:2010gt}%
  \BibitemOpen
  \bibfield  {author} {\bibinfo {author} {\bibfnamefont {J.}~\bibnamefont
  {Fan}}, \bibinfo {author} {\bibfnamefont {M.}~\bibnamefont {Reece}}, \ and\
  \bibinfo {author} {\bibfnamefont {L.-T.}\ \bibnamefont {Wang}},\ }\href
  {\doibase 10.1088/1475-7516/2010/11/042} {\bibfield  {journal} {\bibinfo
  {journal} {JCAP}\ }\textbf {\bibinfo {volume} {1011}},\ \bibinfo {pages}
  {042} (\bibinfo {year} {2010})},\ \Eprint {http://arxiv.org/abs/1008.1591}
  {arXiv:1008.1591 [hep-ph]} \BibitemShut {NoStop}%
\bibitem [{\citenamefont {Fitzpatrick}\ \emph {et~al.}(2013)\citenamefont
  {Fitzpatrick}, \citenamefont {Haxton}, \citenamefont {Katz}, \citenamefont
  {Lubbers},\ and\ \citenamefont {Xu}}]{Fitzpatrick:2012ix}%
  \BibitemOpen
  \bibfield  {author} {\bibinfo {author} {\bibfnamefont {A.~L.}\ \bibnamefont
  {Fitzpatrick}}, \bibinfo {author} {\bibfnamefont {W.}~\bibnamefont {Haxton}},
  \bibinfo {author} {\bibfnamefont {E.}~\bibnamefont {Katz}}, \bibinfo {author}
  {\bibfnamefont {N.}~\bibnamefont {Lubbers}}, \ and\ \bibinfo {author}
  {\bibfnamefont {Y.}~\bibnamefont {Xu}},\ }\href {\doibase
  10.1088/1475-7516/2013/02/004} {\bibfield  {journal} {\bibinfo  {journal}
  {JCAP}\ }\textbf {\bibinfo {volume} {1302}},\ \bibinfo {pages} {004}
  (\bibinfo {year} {2013})},\ \Eprint {http://arxiv.org/abs/1203.3542}
  {arXiv:1203.3542 [hep-ph]} \BibitemShut {NoStop}%
\bibitem [{\citenamefont {Hisano}\ \emph {et~al.}(2005)\citenamefont {Hisano},
  \citenamefont {Matsumoto}, \citenamefont {Nojiri},\ and\ \citenamefont
  {Saito}}]{Hisano:2004ds}%
  \BibitemOpen
  \bibfield  {author} {\bibinfo {author} {\bibfnamefont {J.}~\bibnamefont
  {Hisano}}, \bibinfo {author} {\bibfnamefont {S.}~\bibnamefont {Matsumoto}},
  \bibinfo {author} {\bibfnamefont {M.~M.}\ \bibnamefont {Nojiri}}, \ and\
  \bibinfo {author} {\bibfnamefont {O.}~\bibnamefont {Saito}},\ }\href
  {\doibase 10.1103/PhysRevD.71.063528} {\bibfield  {journal} {\bibinfo
  {journal} {Phys. Rev.}\ }\textbf {\bibinfo {volume} {D71}},\ \bibinfo {pages}
  {063528} (\bibinfo {year} {2005})},\ \Eprint
  {http://arxiv.org/abs/hep-ph/0412403} {arXiv:hep-ph/0412403 [hep-ph]}
  \BibitemShut {NoStop}%
\bibitem [{\citenamefont {Lattanzi}\ and\ \citenamefont
  {Silk}(2009)}]{Lattanzi:2008qa}%
  \BibitemOpen
  \bibfield  {author} {\bibinfo {author} {\bibfnamefont {M.}~\bibnamefont
  {Lattanzi}}\ and\ \bibinfo {author} {\bibfnamefont {J.~I.}\ \bibnamefont
  {Silk}},\ }\href {\doibase 10.1103/PhysRevD.79.083523} {\bibfield  {journal}
  {\bibinfo  {journal} {Phys. Rev.}\ }\textbf {\bibinfo {volume} {D79}},\
  \bibinfo {pages} {083523} (\bibinfo {year} {2009})},\ \Eprint
  {http://arxiv.org/abs/0812.0360} {arXiv:0812.0360 [astro-ph]} \BibitemShut
  {NoStop}%
\bibitem [{\citenamefont {Cirelli}\ \emph {et~al.}(2009)\citenamefont
  {Cirelli}, \citenamefont {Kadastik}, \citenamefont {Raidal},\ and\
  \citenamefont {Strumia}}]{Cirelli:2008pk}%
  \BibitemOpen
  \bibfield  {author} {\bibinfo {author} {\bibfnamefont {M.}~\bibnamefont
  {Cirelli}}, \bibinfo {author} {\bibfnamefont {M.}~\bibnamefont {Kadastik}},
  \bibinfo {author} {\bibfnamefont {M.}~\bibnamefont {Raidal}}, \ and\ \bibinfo
  {author} {\bibfnamefont {A.}~\bibnamefont {Strumia}},\ }\href {\doibase
  10.1016/j.nuclphysb.2013.05.002, 10.1016/j.nuclphysb.2008.11.031} {\bibfield
  {journal} {\bibinfo  {journal} {Nucl. Phys.}\ }\textbf {\bibinfo {volume}
  {B813}},\ \bibinfo {pages} {1} (\bibinfo {year} {2009})},\ \bibinfo {note}
  {[Addendum: Nucl. Phys.B873,530(2013)]},\ \Eprint
  {http://arxiv.org/abs/0809.2409} {arXiv:0809.2409 [hep-ph]} \BibitemShut
  {NoStop}%
\bibitem [{\citenamefont {Arkani-Hamed}\ \emph {et~al.}(2009)\citenamefont
  {Arkani-Hamed}, \citenamefont {Finkbeiner}, \citenamefont {Slatyer},\ and\
  \citenamefont {Weiner}}]{ArkaniHamed:2008qn}%
  \BibitemOpen
  \bibfield  {author} {\bibinfo {author} {\bibfnamefont {N.}~\bibnamefont
  {Arkani-Hamed}}, \bibinfo {author} {\bibfnamefont {D.~P.}\ \bibnamefont
  {Finkbeiner}}, \bibinfo {author} {\bibfnamefont {T.~R.}\ \bibnamefont
  {Slatyer}}, \ and\ \bibinfo {author} {\bibfnamefont {N.}~\bibnamefont
  {Weiner}},\ }\href {\doibase 10.1103/PhysRevD.79.015014} {\bibfield
  {journal} {\bibinfo  {journal} {Phys. Rev.}\ }\textbf {\bibinfo {volume}
  {D79}},\ \bibinfo {pages} {015014} (\bibinfo {year} {2009})},\ \Eprint
  {http://arxiv.org/abs/0810.0713} {arXiv:0810.0713 [hep-ph]} \BibitemShut
  {NoStop}%
\bibitem [{\citenamefont {Pospelov}\ and\ \citenamefont
  {Ritz}(2009)}]{Pospelov:2008jd}%
  \BibitemOpen
  \bibfield  {author} {\bibinfo {author} {\bibfnamefont {M.}~\bibnamefont
  {Pospelov}}\ and\ \bibinfo {author} {\bibfnamefont {A.}~\bibnamefont
  {Ritz}},\ }\href {\doibase 10.1016/j.physletb.2008.12.012} {\bibfield
  {journal} {\bibinfo  {journal} {Phys. Lett.}\ }\textbf {\bibinfo {volume}
  {B671}},\ \bibinfo {pages} {391} (\bibinfo {year} {2009})},\ \Eprint
  {http://arxiv.org/abs/0810.1502} {arXiv:0810.1502 [hep-ph]} \BibitemShut
  {NoStop}%
\bibitem [{\citenamefont {Fox}\ and\ \citenamefont
  {Poppitz}(2009)}]{Fox:2008kb}%
  \BibitemOpen
  \bibfield  {author} {\bibinfo {author} {\bibfnamefont {P.~J.}\ \bibnamefont
  {Fox}}\ and\ \bibinfo {author} {\bibfnamefont {E.}~\bibnamefont {Poppitz}},\
  }\href {\doibase 10.1103/PhysRevD.79.083528} {\bibfield  {journal} {\bibinfo
  {journal} {Phys. Rev.}\ }\textbf {\bibinfo {volume} {D79}},\ \bibinfo {pages}
  {083528} (\bibinfo {year} {2009})},\ \Eprint {http://arxiv.org/abs/0811.0399}
  {arXiv:0811.0399 [hep-ph]} \BibitemShut {NoStop}%
\bibitem [{\citenamefont {Iengo}(2009)}]{Iengo:2009ni}%
  \BibitemOpen
  \bibfield  {author} {\bibinfo {author} {\bibfnamefont {R.}~\bibnamefont
  {Iengo}},\ }\href {\doibase 10.1088/1126-6708/2009/05/024} {\bibfield
  {journal} {\bibinfo  {journal} {JHEP}\ }\textbf {\bibinfo {volume} {05}},\
  \bibinfo {pages} {024} (\bibinfo {year} {2009})},\ \Eprint
  {http://arxiv.org/abs/0902.0688} {arXiv:0902.0688 [hep-ph]} \BibitemShut
  {NoStop}%
\bibitem [{\citenamefont {Hall}\ \emph {et~al.}(2010)\citenamefont {Hall},
  \citenamefont {Jedamzik}, \citenamefont {March-Russell},\ and\ \citenamefont
  {West}}]{Hall:2009bx}%
  \BibitemOpen
  \bibfield  {author} {\bibinfo {author} {\bibfnamefont {L.~J.}\ \bibnamefont
  {Hall}}, \bibinfo {author} {\bibfnamefont {K.}~\bibnamefont {Jedamzik}},
  \bibinfo {author} {\bibfnamefont {J.}~\bibnamefont {March-Russell}}, \ and\
  \bibinfo {author} {\bibfnamefont {S.~M.}\ \bibnamefont {West}},\ }\href
  {\doibase 10.1007/JHEP03(2010)080} {\bibfield  {journal} {\bibinfo  {journal}
  {JHEP}\ }\textbf {\bibinfo {volume} {03}},\ \bibinfo {pages} {080} (\bibinfo
  {year} {2010})},\ \Eprint {http://arxiv.org/abs/0911.1120} {arXiv:0911.1120
  [hep-ph]} \BibitemShut {NoStop}%
\bibitem [{\citenamefont {Co}\ \emph {et~al.}(2015)\citenamefont {Co},
  \citenamefont {D'Eramo}, \citenamefont {Hall},\ and\ \citenamefont
  {Pappadopulo}}]{Co:2015pka}%
  \BibitemOpen
  \bibfield  {author} {\bibinfo {author} {\bibfnamefont {R.~T.}\ \bibnamefont
  {Co}}, \bibinfo {author} {\bibfnamefont {F.}~\bibnamefont {D'Eramo}},
  \bibinfo {author} {\bibfnamefont {L.~J.}\ \bibnamefont {Hall}}, \ and\
  \bibinfo {author} {\bibfnamefont {D.}~\bibnamefont {Pappadopulo}},\ }\href
  {\doibase 10.1088/1475-7516/2015/12/024} {\bibfield  {journal} {\bibinfo
  {journal} {JCAP}\ }\textbf {\bibinfo {volume} {1512}},\ \bibinfo {pages}
  {024} (\bibinfo {year} {2015})},\ \Eprint {http://arxiv.org/abs/1506.07532}
  {arXiv:1506.07532 [hep-ph]} \BibitemShut {NoStop}%
\bibitem [{\citenamefont {Bernal}\ \emph {et~al.}(2017)\citenamefont {Bernal},
  \citenamefont {Heikinheimo}, \citenamefont {Tenkanen}, \citenamefont
  {Tuominen},\ and\ \citenamefont {Vaskonen}}]{Bernal:2017kxu}%
  \BibitemOpen
  \bibfield  {author} {\bibinfo {author} {\bibfnamefont {N.}~\bibnamefont
  {Bernal}}, \bibinfo {author} {\bibfnamefont {M.}~\bibnamefont {Heikinheimo}},
  \bibinfo {author} {\bibfnamefont {T.}~\bibnamefont {Tenkanen}}, \bibinfo
  {author} {\bibfnamefont {K.}~\bibnamefont {Tuominen}}, \ and\ \bibinfo
  {author} {\bibfnamefont {V.}~\bibnamefont {Vaskonen}},\ }\href {\doibase
  10.1142/S0217751X1730023X} {\bibfield  {journal} {\bibinfo  {journal} {Int.
  J. Mod. Phys.}\ }\textbf {\bibinfo {volume} {A32}},\ \bibinfo {pages}
  {1730023} (\bibinfo {year} {2017})},\ \Eprint
  {http://arxiv.org/abs/1706.07442} {arXiv:1706.07442 [hep-ph]} \BibitemShut
  {NoStop}%
\bibitem [{\citenamefont {Kamionkowski}\ and\ \citenamefont
  {Turner}(1990)}]{Kamionkowski:1990ni}%
  \BibitemOpen
  \bibfield  {author} {\bibinfo {author} {\bibfnamefont {M.}~\bibnamefont
  {Kamionkowski}}\ and\ \bibinfo {author} {\bibfnamefont {M.~S.}\ \bibnamefont
  {Turner}},\ }\href {\doibase 10.1103/PhysRevD.42.3310} {\bibfield  {journal}
  {\bibinfo  {journal} {Phys. Rev. D}\ }\textbf {\bibinfo {volume} {42}},\
  \bibinfo {pages} {3310} (\bibinfo {year} {1990})}\BibitemShut {NoStop}%
\bibitem [{\citenamefont {Kofman}\ \emph {et~al.}(1994)\citenamefont {Kofman},
  \citenamefont {Linde},\ and\ \citenamefont {Starobinsky}}]{Kofman:1994rk}%
  \BibitemOpen
  \bibfield  {author} {\bibinfo {author} {\bibfnamefont {L.}~\bibnamefont
  {Kofman}}, \bibinfo {author} {\bibfnamefont {A.~D.}\ \bibnamefont {Linde}}, \
  and\ \bibinfo {author} {\bibfnamefont {A.~A.}\ \bibnamefont {Starobinsky}},\
  }\href {\doibase 10.1103/PhysRevLett.73.3195} {\bibfield  {journal} {\bibinfo
   {journal} {Phys. Rev. Lett.}\ }\textbf {\bibinfo {volume} {73}},\ \bibinfo
  {pages} {3195} (\bibinfo {year} {1994})},\ \Eprint
  {http://arxiv.org/abs/hep-th/9405187} {arXiv:hep-th/9405187 [hep-th]}
  \BibitemShut {NoStop}%
\bibitem [{\citenamefont {Kofman}\ \emph {et~al.}(1997)\citenamefont {Kofman},
  \citenamefont {Linde},\ and\ \citenamefont {Starobinsky}}]{Kofman:1997yn}%
  \BibitemOpen
  \bibfield  {author} {\bibinfo {author} {\bibfnamefont {L.}~\bibnamefont
  {Kofman}}, \bibinfo {author} {\bibfnamefont {A.~D.}\ \bibnamefont {Linde}}, \
  and\ \bibinfo {author} {\bibfnamefont {A.~A.}\ \bibnamefont {Starobinsky}},\
  }\href {\doibase 10.1103/PhysRevD.56.3258} {\bibfield  {journal} {\bibinfo
  {journal} {Phys. Rev.}\ }\textbf {\bibinfo {volume} {D56}},\ \bibinfo {pages}
  {3258} (\bibinfo {year} {1997})},\ \Eprint
  {http://arxiv.org/abs/hep-ph/9704452} {arXiv:hep-ph/9704452 [hep-ph]}
  \BibitemShut {NoStop}%
\bibitem [{\citenamefont {Kawasaki}\ \emph {et~al.}(1999)\citenamefont
  {Kawasaki}, \citenamefont {Kohri},\ and\ \citenamefont
  {Sugiyama}}]{Kawasaki:1999na}%
  \BibitemOpen
  \bibfield  {author} {\bibinfo {author} {\bibfnamefont {M.}~\bibnamefont
  {Kawasaki}}, \bibinfo {author} {\bibfnamefont {K.}~\bibnamefont {Kohri}}, \
  and\ \bibinfo {author} {\bibfnamefont {N.}~\bibnamefont {Sugiyama}},\ }\href
  {\doibase 10.1103/PhysRevLett.82.4168} {\bibfield  {journal} {\bibinfo
  {journal} {Phys. Rev. Lett.}\ }\textbf {\bibinfo {volume} {82}},\ \bibinfo
  {pages} {4168} (\bibinfo {year} {1999})},\ \Eprint
  {http://arxiv.org/abs/astro-ph/9811437} {arXiv:astro-ph/9811437 [astro-ph]}
  \BibitemShut {NoStop}%
\bibitem [{\citenamefont {Kawasaki}\ \emph {et~al.}(2000)\citenamefont
  {Kawasaki}, \citenamefont {Kohri},\ and\ \citenamefont
  {Sugiyama}}]{Kawasaki:2000en}%
  \BibitemOpen
  \bibfield  {author} {\bibinfo {author} {\bibfnamefont {M.}~\bibnamefont
  {Kawasaki}}, \bibinfo {author} {\bibfnamefont {K.}~\bibnamefont {Kohri}}, \
  and\ \bibinfo {author} {\bibfnamefont {N.}~\bibnamefont {Sugiyama}},\ }\href
  {\doibase 10.1103/PhysRevD.62.023506} {\bibfield  {journal} {\bibinfo
  {journal} {Phys. Rev.}\ }\textbf {\bibinfo {volume} {D62}},\ \bibinfo {pages}
  {023506} (\bibinfo {year} {2000})},\ \Eprint
  {http://arxiv.org/abs/astro-ph/0002127} {arXiv:astro-ph/0002127 [astro-ph]}
  \BibitemShut {NoStop}%
\bibitem [{\citenamefont {Hannestad}(2004)}]{Hannestad:2004px}%
  \BibitemOpen
  \bibfield  {author} {\bibinfo {author} {\bibfnamefont {S.}~\bibnamefont
  {Hannestad}},\ }\href {\doibase 10.1103/PhysRevD.70.043506} {\bibfield
  {journal} {\bibinfo  {journal} {Phys. Rev.}\ }\textbf {\bibinfo {volume}
  {D70}},\ \bibinfo {pages} {043506} (\bibinfo {year} {2004})},\ \Eprint
  {http://arxiv.org/abs/astro-ph/0403291} {arXiv:astro-ph/0403291 [astro-ph]}
  \BibitemShut {NoStop}%
\bibitem [{\citenamefont {Ichikawa}\ \emph {et~al.}(2005)\citenamefont
  {Ichikawa}, \citenamefont {Kawasaki},\ and\ \citenamefont
  {Takahashi}}]{Ichikawa:2005vw}%
  \BibitemOpen
  \bibfield  {author} {\bibinfo {author} {\bibfnamefont {K.}~\bibnamefont
  {Ichikawa}}, \bibinfo {author} {\bibfnamefont {M.}~\bibnamefont {Kawasaki}},
  \ and\ \bibinfo {author} {\bibfnamefont {F.}~\bibnamefont {Takahashi}},\
  }\href {\doibase 10.1103/PhysRevD.72.043522} {\bibfield  {journal} {\bibinfo
  {journal} {Phys. Rev.}\ }\textbf {\bibinfo {volume} {D72}},\ \bibinfo {pages}
  {043522} (\bibinfo {year} {2005})},\ \Eprint
  {http://arxiv.org/abs/astro-ph/0505395} {arXiv:astro-ph/0505395 [astro-ph]}
  \BibitemShut {NoStop}%
\bibitem [{\citenamefont {De~Bernardis}\ \emph {et~al.}(2008)\citenamefont
  {De~Bernardis}, \citenamefont {Pagano},\ and\ \citenamefont
  {Melchiorri}}]{DeBernardis:2008zz}%
  \BibitemOpen
  \bibfield  {author} {\bibinfo {author} {\bibfnamefont {F.}~\bibnamefont
  {De~Bernardis}}, \bibinfo {author} {\bibfnamefont {L.}~\bibnamefont
  {Pagano}}, \ and\ \bibinfo {author} {\bibfnamefont {A.}~\bibnamefont
  {Melchiorri}},\ }\href {\doibase 10.1016/j.astropartphys.2008.09.005}
  {\bibfield  {journal} {\bibinfo  {journal} {Astropart. Phys.}\ }\textbf
  {\bibinfo {volume} {30}},\ \bibinfo {pages} {192} (\bibinfo {year}
  {2008})}\BibitemShut {NoStop}%
\bibitem [{\citenamefont {Gelmini}\ and\ \citenamefont
  {Gondolo}(2008)}]{Gelmini:2008sh}%
  \BibitemOpen
  \bibfield  {author} {\bibinfo {author} {\bibfnamefont {G.~B.}\ \bibnamefont
  {Gelmini}}\ and\ \bibinfo {author} {\bibfnamefont {P.}~\bibnamefont
  {Gondolo}},\ }\href {\doibase 10.1088/1475-7516/2008/10/002} {\bibfield
  {journal} {\bibinfo  {journal} {JCAP}\ }\textbf {\bibinfo {volume} {0810}},\
  \bibinfo {pages} {002} (\bibinfo {year} {2008})},\ \Eprint
  {http://arxiv.org/abs/0803.2349} {arXiv:0803.2349 [astro-ph]} \BibitemShut
  {NoStop}%
\bibitem [{\citenamefont {Visinelli}\ and\ \citenamefont
  {Gondolo}(2015)}]{Visinelli:2015eka}%
  \BibitemOpen
  \bibfield  {author} {\bibinfo {author} {\bibfnamefont {L.}~\bibnamefont
  {Visinelli}}\ and\ \bibinfo {author} {\bibfnamefont {P.}~\bibnamefont
  {Gondolo}},\ }\href {\doibase 10.1103/PhysRevD.91.083526} {\bibfield
  {journal} {\bibinfo  {journal} {Phys. Rev.}\ }\textbf {\bibinfo {volume}
  {D91}},\ \bibinfo {pages} {083526} (\bibinfo {year} {2015})},\ \Eprint
  {http://arxiv.org/abs/1501.02233} {arXiv:1501.02233 [astro-ph.CO]}
  \BibitemShut {NoStop}%
\bibitem [{\citenamefont {Waldstein}\ \emph {et~al.}(2017)\citenamefont
  {Waldstein}, \citenamefont {Erickcek},\ and\ \citenamefont
  {Ilie}}]{Waldstein:2016blt}%
  \BibitemOpen
  \bibfield  {author} {\bibinfo {author} {\bibfnamefont {I.~R.}\ \bibnamefont
  {Waldstein}}, \bibinfo {author} {\bibfnamefont {A.~L.}\ \bibnamefont
  {Erickcek}}, \ and\ \bibinfo {author} {\bibfnamefont {C.}~\bibnamefont
  {Ilie}},\ }\href {\doibase 10.1103/PhysRevD.95.123531} {\bibfield  {journal}
  {\bibinfo  {journal} {Phys. Rev.}\ }\textbf {\bibinfo {volume} {D95}},\
  \bibinfo {pages} {123531} (\bibinfo {year} {2017})},\ \Eprint
  {http://arxiv.org/abs/1609.05927} {arXiv:1609.05927 [astro-ph.CO]}
  \BibitemShut {NoStop}%
\bibitem [{\citenamefont {Waldstein}\ and\ \citenamefont
  {Erickcek}(2017)}]{Waldstein:2017wps}%
  \BibitemOpen
  \bibfield  {author} {\bibinfo {author} {\bibfnamefont {I.~R.}\ \bibnamefont
  {Waldstein}}\ and\ \bibinfo {author} {\bibfnamefont {A.~L.}\ \bibnamefont
  {Erickcek}},\ }\href {\doibase 10.1103/PhysRevD.95.088301} {\bibfield
  {journal} {\bibinfo  {journal} {Phys. Rev.}\ }\textbf {\bibinfo {volume}
  {D95}},\ \bibinfo {pages} {088301} (\bibinfo {year} {2017})},\ \Eprint
  {http://arxiv.org/abs/1707.03417} {arXiv:1707.03417 [astro-ph.CO]}
  \BibitemShut {NoStop}%
\bibitem [{\citenamefont {Dine}\ and\ \citenamefont
  {Fischler}(1983)}]{Dine:1982ah}%
  \BibitemOpen
  \bibfield  {author} {\bibinfo {author} {\bibfnamefont {M.}~\bibnamefont
  {Dine}}\ and\ \bibinfo {author} {\bibfnamefont {W.}~\bibnamefont
  {Fischler}},\ }\href {\doibase 10.1016/0370-2693(83)90639-1} {\bibfield
  {journal} {\bibinfo  {journal} {Phys. Lett.}\ }\textbf {\bibinfo {volume}
  {120B}},\ \bibinfo {pages} {137} (\bibinfo {year} {1983})}\BibitemShut
  {NoStop}%
\bibitem [{\citenamefont {Steinhardt}\ and\ \citenamefont
  {Turner}(1983)}]{Steinhardt:1983ia}%
  \BibitemOpen
  \bibfield  {author} {\bibinfo {author} {\bibfnamefont {P.~J.}\ \bibnamefont
  {Steinhardt}}\ and\ \bibinfo {author} {\bibfnamefont {M.~S.}\ \bibnamefont
  {Turner}},\ }\href {\doibase 10.1016/0370-2693(83)90727-X} {\bibfield
  {journal} {\bibinfo  {journal} {Phys. Lett.}\ }\textbf {\bibinfo {volume}
  {129B}},\ \bibinfo {pages} {51} (\bibinfo {year} {1983})}\BibitemShut
  {NoStop}%
\bibitem [{\citenamefont {Turner}(1983)}]{Turner:1983he}%
  \BibitemOpen
  \bibfield  {author} {\bibinfo {author} {\bibfnamefont {M.~S.}\ \bibnamefont
  {Turner}},\ }\href {\doibase 10.1103/PhysRevD.28.1243} {\bibfield  {journal}
  {\bibinfo  {journal} {Phys. Rev.}\ }\textbf {\bibinfo {volume} {D28}},\
  \bibinfo {pages} {1243} (\bibinfo {year} {1983})}\BibitemShut {NoStop}%
\bibitem [{\citenamefont {Scherrer}\ and\ \citenamefont
  {Turner}(1985)}]{Scherrer:1984fd}%
  \BibitemOpen
  \bibfield  {author} {\bibinfo {author} {\bibfnamefont {R.~J.}\ \bibnamefont
  {Scherrer}}\ and\ \bibinfo {author} {\bibfnamefont {M.~S.}\ \bibnamefont
  {Turner}},\ }\href {\doibase 10.1103/PhysRevD.31.681} {\bibfield  {journal}
  {\bibinfo  {journal} {Phys. Rev.}\ }\textbf {\bibinfo {volume} {D31}},\
  \bibinfo {pages} {681} (\bibinfo {year} {1985})}\BibitemShut {NoStop}%
\bibitem [{\citenamefont {Lyth}\ and\ \citenamefont
  {Stewart}(1996)}]{Lyth:1995ka}%
  \BibitemOpen
  \bibfield  {author} {\bibinfo {author} {\bibfnamefont {D.~H.}\ \bibnamefont
  {Lyth}}\ and\ \bibinfo {author} {\bibfnamefont {E.~D.}\ \bibnamefont
  {Stewart}},\ }\href {\doibase 10.1103/PhysRevD.53.1784} {\bibfield  {journal}
  {\bibinfo  {journal} {Phys. Rev.}\ }\textbf {\bibinfo {volume} {D53}},\
  \bibinfo {pages} {1784} (\bibinfo {year} {1996})},\ \Eprint
  {http://arxiv.org/abs/hep-ph/9510204} {arXiv:hep-ph/9510204 [hep-ph]}
  \BibitemShut {NoStop}%
\bibitem [{\citenamefont {Chung}\ \emph {et~al.}(1999)\citenamefont {Chung},
  \citenamefont {Kolb},\ and\ \citenamefont {Riotto}}]{Chung:1998rq}%
  \BibitemOpen
  \bibfield  {author} {\bibinfo {author} {\bibfnamefont {D.~J.~H.}\
  \bibnamefont {Chung}}, \bibinfo {author} {\bibfnamefont {E.~W.}\ \bibnamefont
  {Kolb}}, \ and\ \bibinfo {author} {\bibfnamefont {A.}~\bibnamefont
  {Riotto}},\ }\href {\doibase 10.1103/PhysRevD.60.063504} {\bibfield
  {journal} {\bibinfo  {journal} {Phys. Rev.}\ }\textbf {\bibinfo {volume}
  {D60}},\ \bibinfo {pages} {063504} (\bibinfo {year} {1999})},\ \Eprint
  {http://arxiv.org/abs/hep-ph/9809453} {arXiv:hep-ph/9809453 [hep-ph]}
  \BibitemShut {NoStop}%
\bibitem [{\citenamefont {Giudice}\ \emph {et~al.}(2001)\citenamefont
  {Giudice}, \citenamefont {Kolb},\ and\ \citenamefont
  {Riotto}}]{Giudice:2000ex}%
  \BibitemOpen
  \bibfield  {author} {\bibinfo {author} {\bibfnamefont {G.~F.}\ \bibnamefont
  {Giudice}}, \bibinfo {author} {\bibfnamefont {E.~W.}\ \bibnamefont {Kolb}}, \
  and\ \bibinfo {author} {\bibfnamefont {A.}~\bibnamefont {Riotto}},\ }\href
  {\doibase 10.1103/PhysRevD.64.023508} {\bibfield  {journal} {\bibinfo
  {journal} {Phys. Rev.}\ }\textbf {\bibinfo {volume} {D64}},\ \bibinfo {pages}
  {023508} (\bibinfo {year} {2001})},\ \Eprint
  {http://arxiv.org/abs/hep-ph/0005123} {arXiv:hep-ph/0005123 [hep-ph]}
  \BibitemShut {NoStop}%
\bibitem [{\citenamefont {{Moroi}}\ and\ \citenamefont
  {{Randall}}(2000)}]{Moroi:2000}%
  \BibitemOpen
  \bibfield  {author} {\bibinfo {author} {\bibfnamefont {T.}~\bibnamefont
  {{Moroi}}}\ and\ \bibinfo {author} {\bibfnamefont {L.}~\bibnamefont
  {{Randall}}},\ }\href {\doibase 10.1016/S0550-3213(99)00748-8} {\bibfield
  {journal} {\bibinfo  {journal} {Nuclear Physics B}\ }\textbf {\bibinfo
  {volume} {570}},\ \bibinfo {pages} {455} (\bibinfo {year} {2000})},\ \Eprint
  {http://arxiv.org/abs/hep-ph/9906527} {hep-ph/9906527} \BibitemShut {NoStop}%
\bibitem [{\citenamefont {Fujii}\ and\ \citenamefont
  {Hamaguchi}(2002)}]{Fujii:2002kr}%
  \BibitemOpen
  \bibfield  {author} {\bibinfo {author} {\bibfnamefont {M.}~\bibnamefont
  {Fujii}}\ and\ \bibinfo {author} {\bibfnamefont {K.}~\bibnamefont
  {Hamaguchi}},\ }\href {\doibase 10.1103/PhysRevD.66.083501} {\bibfield
  {journal} {\bibinfo  {journal} {Phys. Rev.}\ }\textbf {\bibinfo {volume}
  {D66}},\ \bibinfo {pages} {083501} (\bibinfo {year} {2002})},\ \Eprint
  {http://arxiv.org/abs/hep-ph/0205044} {arXiv:hep-ph/0205044 [hep-ph]}
  \BibitemShut {NoStop}%
\bibitem [{\citenamefont {Fujii}\ \emph {et~al.}(2004)\citenamefont {Fujii},
  \citenamefont {Ibe},\ and\ \citenamefont {Yanagida}}]{Fujii:2003iw}%
  \BibitemOpen
  \bibfield  {author} {\bibinfo {author} {\bibfnamefont {M.}~\bibnamefont
  {Fujii}}, \bibinfo {author} {\bibfnamefont {M.}~\bibnamefont {Ibe}}, \ and\
  \bibinfo {author} {\bibfnamefont {T.}~\bibnamefont {Yanagida}},\ }\href
  {\doibase 10.1103/PhysRevD.69.015006} {\bibfield  {journal} {\bibinfo
  {journal} {Phys. Rev.}\ }\textbf {\bibinfo {volume} {D69}},\ \bibinfo {pages}
  {015006} (\bibinfo {year} {2004})},\ \Eprint
  {http://arxiv.org/abs/hep-ph/0309064} {arXiv:hep-ph/0309064 [hep-ph]}
  \BibitemShut {NoStop}%
\bibitem [{\citenamefont {Gelmini}\ and\ \citenamefont
  {Gondolo}(2006)}]{Gelmini:2006pw}%
  \BibitemOpen
  \bibfield  {author} {\bibinfo {author} {\bibfnamefont {G.~B.}\ \bibnamefont
  {Gelmini}}\ and\ \bibinfo {author} {\bibfnamefont {P.}~\bibnamefont
  {Gondolo}},\ }\href {\doibase 10.1103/PhysRevD.74.023510} {\bibfield
  {journal} {\bibinfo  {journal} {Phys. Rev.}\ }\textbf {\bibinfo {volume}
  {D74}},\ \bibinfo {pages} {023510} (\bibinfo {year} {2006})},\ \Eprint
  {http://arxiv.org/abs/hep-ph/0602230} {arXiv:hep-ph/0602230 [hep-ph]}
  \BibitemShut {NoStop}%
\bibitem [{\citenamefont {Gelmini}\ \emph {et~al.}(2006)\citenamefont
  {Gelmini}, \citenamefont {Gondolo}, \citenamefont {Soldatenko},\ and\
  \citenamefont {Yaguna}}]{Gelmini:2006pq}%
  \BibitemOpen
  \bibfield  {author} {\bibinfo {author} {\bibfnamefont {G.}~\bibnamefont
  {Gelmini}}, \bibinfo {author} {\bibfnamefont {P.}~\bibnamefont {Gondolo}},
  \bibinfo {author} {\bibfnamefont {A.}~\bibnamefont {Soldatenko}}, \ and\
  \bibinfo {author} {\bibfnamefont {C.~E.}\ \bibnamefont {Yaguna}},\ }\href
  {\doibase 10.1103/PhysRevD.74.083514} {\bibfield  {journal} {\bibinfo
  {journal} {Phys. Rev.}\ }\textbf {\bibinfo {volume} {D74}},\ \bibinfo {pages}
  {083514} (\bibinfo {year} {2006})},\ \Eprint
  {http://arxiv.org/abs/hep-ph/0605016} {arXiv:hep-ph/0605016 [hep-ph]}
  \BibitemShut {NoStop}%
\bibitem [{\citenamefont {Acharya}\ \emph {et~al.}(2009)\citenamefont
  {Acharya}, \citenamefont {Kane}, \citenamefont {Watson},\ and\ \citenamefont
  {Kumar}}]{Acharya:2009zt}%
  \BibitemOpen
  \bibfield  {author} {\bibinfo {author} {\bibfnamefont {B.~S.}\ \bibnamefont
  {Acharya}}, \bibinfo {author} {\bibfnamefont {G.}~\bibnamefont {Kane}},
  \bibinfo {author} {\bibfnamefont {S.}~\bibnamefont {Watson}}, \ and\ \bibinfo
  {author} {\bibfnamefont {P.}~\bibnamefont {Kumar}},\ }\href {\doibase
  10.1103/PhysRevD.80.083529} {\bibfield  {journal} {\bibinfo  {journal} {Phys.
  Rev.}\ }\textbf {\bibinfo {volume} {D80}},\ \bibinfo {pages} {083529}
  (\bibinfo {year} {2009})},\ \Eprint {http://arxiv.org/abs/0908.2430}
  {arXiv:0908.2430 [astro-ph.CO]} \BibitemShut {NoStop}%
\bibitem [{\citenamefont {Grin}\ \emph {et~al.}(2010)\citenamefont {Grin},
  \citenamefont {Smith},\ and\ \citenamefont {Kamionkowski}}]{Grin:2010zz}%
  \BibitemOpen
  \bibfield  {author} {\bibinfo {author} {\bibfnamefont {D.}~\bibnamefont
  {Grin}}, \bibinfo {author} {\bibfnamefont {T.}~\bibnamefont {Smith}}, \ and\
  \bibinfo {author} {\bibfnamefont {M.}~\bibnamefont {Kamionkowski}},\
  }\bibfield  {booktitle} {\emph {\bibinfo {booktitle} {{Proceedings,
  International Conference on Axions 2010: Gainesville, Florida, January 15-17,
  2010}}},\ }\href {\doibase 10.1063/1.3489561} {\bibfield  {journal} {\bibinfo
   {journal} {AIP Conf. Proc.}\ }\textbf {\bibinfo {volume} {1274}},\ \bibinfo
  {pages} {78} (\bibinfo {year} {2010})}\BibitemShut {NoStop}%
\bibitem [{\citenamefont {Harigaya}\ \emph {et~al.}(2014)\citenamefont
  {Harigaya}, \citenamefont {Kawasaki}, \citenamefont {Mukaida},\ and\
  \citenamefont {Yamada}}]{Harigaya:2014waa}%
  \BibitemOpen
  \bibfield  {author} {\bibinfo {author} {\bibfnamefont {K.}~\bibnamefont
  {Harigaya}}, \bibinfo {author} {\bibfnamefont {M.}~\bibnamefont {Kawasaki}},
  \bibinfo {author} {\bibfnamefont {K.}~\bibnamefont {Mukaida}}, \ and\
  \bibinfo {author} {\bibfnamefont {M.}~\bibnamefont {Yamada}},\ }\href
  {\doibase 10.1103/PhysRevD.89.083532} {\bibfield  {journal} {\bibinfo
  {journal} {Phys. Rev.}\ }\textbf {\bibinfo {volume} {D89}},\ \bibinfo {pages}
  {083532} (\bibinfo {year} {2014})},\ \Eprint {http://arxiv.org/abs/1402.2846}
  {arXiv:1402.2846 [hep-ph]} \BibitemShut {NoStop}%
\bibitem [{\citenamefont {Baer}\ \emph {et~al.}(2015)\citenamefont {Baer},
  \citenamefont {Choi}, \citenamefont {Kim},\ and\ \citenamefont
  {Roszkowski}}]{Baer:2014eja}%
  \BibitemOpen
  \bibfield  {author} {\bibinfo {author} {\bibfnamefont {H.}~\bibnamefont
  {Baer}}, \bibinfo {author} {\bibfnamefont {K.-Y.}\ \bibnamefont {Choi}},
  \bibinfo {author} {\bibfnamefont {J.~E.}\ \bibnamefont {Kim}}, \ and\
  \bibinfo {author} {\bibfnamefont {L.}~\bibnamefont {Roszkowski}},\ }\href
  {\doibase 10.1016/j.physrep.2014.10.002} {\bibfield  {journal} {\bibinfo
  {journal} {Phys. Rept.}\ }\textbf {\bibinfo {volume} {555}},\ \bibinfo
  {pages} {1} (\bibinfo {year} {2015})},\ \Eprint
  {http://arxiv.org/abs/1407.0017} {arXiv:1407.0017 [hep-ph]} \BibitemShut
  {NoStop}%
\bibitem [{\citenamefont {Monteux}\ and\ \citenamefont
  {Shin}(2015)}]{Monteux:2015qqa}%
  \BibitemOpen
  \bibfield  {author} {\bibinfo {author} {\bibfnamefont {A.}~\bibnamefont
  {Monteux}}\ and\ \bibinfo {author} {\bibfnamefont {C.~S.}\ \bibnamefont
  {Shin}},\ }\href {\doibase 10.1103/PhysRevD.92.035002} {\bibfield  {journal}
  {\bibinfo  {journal} {Phys. Rev.}\ }\textbf {\bibinfo {volume} {D92}},\
  \bibinfo {pages} {035002} (\bibinfo {year} {2015})},\ \Eprint
  {http://arxiv.org/abs/1505.03149} {arXiv:1505.03149 [hep-ph]} \BibitemShut
  {NoStop}%
\bibitem [{\citenamefont {Reece}\ and\ \citenamefont
  {Roxlo}(2016)}]{Reece:2015lch}%
  \BibitemOpen
  \bibfield  {author} {\bibinfo {author} {\bibfnamefont {M.}~\bibnamefont
  {Reece}}\ and\ \bibinfo {author} {\bibfnamefont {T.}~\bibnamefont {Roxlo}},\
  }\href {\doibase 10.1007/JHEP09(2016)096} {\bibfield  {journal} {\bibinfo
  {journal} {JHEP}\ }\textbf {\bibinfo {volume} {09}},\ \bibinfo {pages} {096}
  (\bibinfo {year} {2016})},\ \Eprint {http://arxiv.org/abs/1511.06768}
  {arXiv:1511.06768 [hep-ph]} \BibitemShut {NoStop}%
\bibitem [{\citenamefont {Kane}\ \emph {et~al.}(2016)\citenamefont {Kane},
  \citenamefont {Kumar}, \citenamefont {Nelson},\ and\ \citenamefont
  {Zheng}}]{Kane:2015qea}%
  \BibitemOpen
  \bibfield  {author} {\bibinfo {author} {\bibfnamefont {G.~L.}\ \bibnamefont
  {Kane}}, \bibinfo {author} {\bibfnamefont {P.}~\bibnamefont {Kumar}},
  \bibinfo {author} {\bibfnamefont {B.~D.}\ \bibnamefont {Nelson}}, \ and\
  \bibinfo {author} {\bibfnamefont {B.}~\bibnamefont {Zheng}},\ }\href
  {\doibase 10.1103/PhysRevD.93.063527} {\bibfield  {journal} {\bibinfo
  {journal} {Phys. Rev.}\ }\textbf {\bibinfo {volume} {D93}},\ \bibinfo {pages}
  {063527} (\bibinfo {year} {2016})},\ \Eprint
  {http://arxiv.org/abs/1502.05406} {arXiv:1502.05406 [hep-ph]} \BibitemShut
  {NoStop}%
\bibitem [{\citenamefont {Erickcek}(2015)}]{Erickcek:2015jza}%
  \BibitemOpen
  \bibfield  {author} {\bibinfo {author} {\bibfnamefont {A.~L.}\ \bibnamefont
  {Erickcek}},\ }\href {\doibase 10.1103/PhysRevD.92.103505} {\bibfield
  {journal} {\bibinfo  {journal} {Phys. Rev.}\ }\textbf {\bibinfo {volume}
  {D92}},\ \bibinfo {pages} {103505} (\bibinfo {year} {2015})},\ \Eprint
  {http://arxiv.org/abs/1504.03335} {arXiv:1504.03335 [astro-ph.CO]}
  \BibitemShut {NoStop}%
\bibitem [{\citenamefont {Kim}\ \emph {et~al.}(2017)\citenamefont {Kim},
  \citenamefont {Hong},\ and\ \citenamefont {Shin}}]{Kim:2016spf}%
  \BibitemOpen
  \bibfield  {author} {\bibinfo {author} {\bibfnamefont {H.}~\bibnamefont
  {Kim}}, \bibinfo {author} {\bibfnamefont {J.-P.}\ \bibnamefont {Hong}}, \
  and\ \bibinfo {author} {\bibfnamefont {C.~S.}\ \bibnamefont {Shin}},\ }\href
  {\doibase 10.1016/j.physletb.2017.03.005} {\bibfield  {journal} {\bibinfo
  {journal} {Phys. Lett.}\ }\textbf {\bibinfo {volume} {B768}},\ \bibinfo
  {pages} {292} (\bibinfo {year} {2017})},\ \Eprint
  {http://arxiv.org/abs/1611.02287} {arXiv:1611.02287 [hep-ph]} \BibitemShut
  {NoStop}%
\bibitem [{\citenamefont {Barrow}(1982)}]{Barrow:1982}%
  \BibitemOpen
  \bibfield  {author} {\bibinfo {author} {\bibfnamefont {J.~D.}\ \bibnamefont
  {Barrow}},\ }\href@noop {} {\bibfield  {journal} {\bibinfo  {journal} {Nucl.
  Phys.}\ }\textbf {\bibinfo {volume} {B208}},\ \bibinfo {pages} {501}
  (\bibinfo {year} {1982})}\BibitemShut {NoStop}%
\bibitem [{\citenamefont {Ford}(1987)}]{Ford:1986sy}%
  \BibitemOpen
  \bibfield  {author} {\bibinfo {author} {\bibfnamefont {L.~H.}\ \bibnamefont
  {Ford}},\ }\href {\doibase 10.1103/PhysRevD.35.2955} {\bibfield  {journal}
  {\bibinfo  {journal} {Phys. Rev.}\ }\textbf {\bibinfo {volume} {D35}},\
  \bibinfo {pages} {2955} (\bibinfo {year} {1987})}\BibitemShut {NoStop}%
\bibitem [{\citenamefont {Spokoiny}(1993)}]{Spokoiny:1993kt}%
  \BibitemOpen
  \bibfield  {author} {\bibinfo {author} {\bibfnamefont {B.}~\bibnamefont
  {Spokoiny}},\ }\href {\doibase 10.1016/0370-2693(93)90155-B} {\bibfield
  {journal} {\bibinfo  {journal} {Phys. Lett.}\ }\textbf {\bibinfo {volume}
  {B315}},\ \bibinfo {pages} {40} (\bibinfo {year} {1993})},\ \Eprint
  {http://arxiv.org/abs/gr-qc/9306008} {arXiv:gr-qc/9306008 [gr-qc]}
  \BibitemShut {NoStop}%
\bibitem [{\citenamefont {Joyce}(1997)}]{Joyce:1996cp}%
  \BibitemOpen
  \bibfield  {author} {\bibinfo {author} {\bibfnamefont {M.}~\bibnamefont
  {Joyce}},\ }\href {\doibase 10.1103/PhysRevD.55.1875} {\bibfield  {journal}
  {\bibinfo  {journal} {Phys. Rev.}\ }\textbf {\bibinfo {volume} {D55}},\
  \bibinfo {pages} {1875} (\bibinfo {year} {1997})},\ \Eprint
  {http://arxiv.org/abs/hep-ph/9606223} {arXiv:hep-ph/9606223 [hep-ph]}
  \BibitemShut {NoStop}%
\bibitem [{\citenamefont {Salati}(2003)}]{Salati:2002md}%
  \BibitemOpen
  \bibfield  {author} {\bibinfo {author} {\bibfnamefont {P.}~\bibnamefont
  {Salati}},\ }\href {\doibase 10.1016/j.physletb.2003.07.073} {\bibfield
  {journal} {\bibinfo  {journal} {Phys. Lett.}\ }\textbf {\bibinfo {volume}
  {B571}},\ \bibinfo {pages} {121} (\bibinfo {year} {2003})},\ \Eprint
  {http://arxiv.org/abs/astro-ph/0207396} {arXiv:astro-ph/0207396 [astro-ph]}
  \BibitemShut {NoStop}%
\bibitem [{\citenamefont {Profumo}\ and\ \citenamefont
  {Ullio}(2003)}]{Profumo:2003hq}%
  \BibitemOpen
  \bibfield  {author} {\bibinfo {author} {\bibfnamefont {S.}~\bibnamefont
  {Profumo}}\ and\ \bibinfo {author} {\bibfnamefont {P.}~\bibnamefont
  {Ullio}},\ }\href {\doibase 10.1088/1475-7516/2003/11/006} {\bibfield
  {journal} {\bibinfo  {journal} {JCAP}\ }\textbf {\bibinfo {volume} {0311}},\
  \bibinfo {pages} {006} (\bibinfo {year} {2003})},\ \Eprint
  {http://arxiv.org/abs/hep-ph/0309220} {arXiv:hep-ph/0309220 [hep-ph]}
  \BibitemShut {NoStop}%
\bibitem [{\citenamefont {Feng}\ \emph {et~al.}(2003)\citenamefont {Feng},
  \citenamefont {Rajaraman},\ and\ \citenamefont {Takayama}}]{Feng:2003uy}%
  \BibitemOpen
  \bibfield  {author} {\bibinfo {author} {\bibfnamefont {J.~L.}\ \bibnamefont
  {Feng}}, \bibinfo {author} {\bibfnamefont {A.}~\bibnamefont {Rajaraman}}, \
  and\ \bibinfo {author} {\bibfnamefont {F.}~\bibnamefont {Takayama}},\ }\href
  {\doibase 10.1103/PhysRevD.68.063504} {\bibfield  {journal} {\bibinfo
  {journal} {Phys. Rev.}\ }\textbf {\bibinfo {volume} {D68}},\ \bibinfo {pages}
  {063504} (\bibinfo {year} {2003})},\ \Eprint
  {http://arxiv.org/abs/hep-ph/0306024} {arXiv:hep-ph/0306024 [hep-ph]}
  \BibitemShut {NoStop}%
\bibitem [{\citenamefont {Feng}\ \emph {et~al.}(2004)\citenamefont {Feng},
  \citenamefont {Su},\ and\ \citenamefont {Takayama}}]{Feng:2004zu}%
  \BibitemOpen
  \bibfield  {author} {\bibinfo {author} {\bibfnamefont {J.~L.}\ \bibnamefont
  {Feng}}, \bibinfo {author} {\bibfnamefont {S.-f.}\ \bibnamefont {Su}}, \ and\
  \bibinfo {author} {\bibfnamefont {F.}~\bibnamefont {Takayama}},\ }\href
  {\doibase 10.1103/PhysRevD.70.063514} {\bibfield  {journal} {\bibinfo
  {journal} {Phys. Rev.}\ }\textbf {\bibinfo {volume} {D70}},\ \bibinfo {pages}
  {063514} (\bibinfo {year} {2004})},\ \Eprint
  {http://arxiv.org/abs/hep-ph/0404198} {arXiv:hep-ph/0404198 [hep-ph]}
  \BibitemShut {NoStop}%
\bibitem [{\citenamefont {Pallis}(2005)}]{Pallis:2005hm}%
  \BibitemOpen
  \bibfield  {author} {\bibinfo {author} {\bibfnamefont {C.}~\bibnamefont
  {Pallis}},\ }\href {\doibase 10.1088/1475-7516/2005/10/015} {\bibfield
  {journal} {\bibinfo  {journal} {JCAP}\ }\textbf {\bibinfo {volume} {0510}},\
  \bibinfo {pages} {015} (\bibinfo {year} {2005})},\ \Eprint
  {http://arxiv.org/abs/hep-ph/0503080} {arXiv:hep-ph/0503080 [hep-ph]}
  \BibitemShut {NoStop}%
\bibitem [{\citenamefont {Gomez}\ \emph {et~al.}(2009)\citenamefont {Gomez},
  \citenamefont {Lola}, \citenamefont {Pallis},\ and\ \citenamefont
  {Rodriguez-Quintero}}]{Gomez:2008qi}%
  \BibitemOpen
  \bibfield  {author} {\bibinfo {author} {\bibfnamefont {M.~E.}\ \bibnamefont
  {Gomez}}, \bibinfo {author} {\bibfnamefont {S.}~\bibnamefont {Lola}},
  \bibinfo {author} {\bibfnamefont {C.}~\bibnamefont {Pallis}}, \ and\ \bibinfo
  {author} {\bibfnamefont {J.}~\bibnamefont {Rodriguez-Quintero}},\ }\bibfield
  {booktitle} {\emph {\bibinfo {booktitle} {{Proceedings, 4th International
  Workshop on the Dark Side of the Universe (DSU 2008): Cairo, Egypt, June 1-5,
  2008}}},\ }\href {\doibase 10.1063/1.3131490} {\bibfield  {journal} {\bibinfo
   {journal} {AIP Conf. Proc.}\ }\textbf {\bibinfo {volume} {1115}},\ \bibinfo
  {pages} {157} (\bibinfo {year} {2009})},\ \Eprint
  {http://arxiv.org/abs/0809.1982} {arXiv:0809.1982 [hep-ph]} \BibitemShut
  {NoStop}%
\bibitem [{\citenamefont {Lola}\ \emph {et~al.}(2009)\citenamefont {Lola},
  \citenamefont {Pallis},\ and\ \citenamefont {Tzelati}}]{Lola:2009at}%
  \BibitemOpen
  \bibfield  {author} {\bibinfo {author} {\bibfnamefont {S.}~\bibnamefont
  {Lola}}, \bibinfo {author} {\bibfnamefont {C.}~\bibnamefont {Pallis}}, \ and\
  \bibinfo {author} {\bibfnamefont {E.}~\bibnamefont {Tzelati}},\ }\href
  {\doibase 10.1088/1475-7516/2009/11/017} {\bibfield  {journal} {\bibinfo
  {journal} {JCAP}\ }\textbf {\bibinfo {volume} {0911}},\ \bibinfo {pages}
  {017} (\bibinfo {year} {2009})},\ \Eprint {http://arxiv.org/abs/0907.2941}
  {arXiv:0907.2941 [hep-ph]} \BibitemShut {NoStop}%
\bibitem [{\citenamefont {Lewicki}\ \emph {et~al.}(2016)\citenamefont
  {Lewicki}, \citenamefont {Rindler-Daller},\ and\ \citenamefont
  {Wells}}]{Lewicki:2016efe}%
  \BibitemOpen
  \bibfield  {author} {\bibinfo {author} {\bibfnamefont {M.}~\bibnamefont
  {Lewicki}}, \bibinfo {author} {\bibfnamefont {T.}~\bibnamefont
  {Rindler-Daller}}, \ and\ \bibinfo {author} {\bibfnamefont {J.~D.}\
  \bibnamefont {Wells}},\ }\href {\doibase 10.1007/JHEP06(2016)055} {\bibfield
  {journal} {\bibinfo  {journal} {JHEP}\ }\textbf {\bibinfo {volume} {06}},\
  \bibinfo {pages} {055} (\bibinfo {year} {2016})},\ \Eprint
  {http://arxiv.org/abs/1601.01681} {arXiv:1601.01681 [hep-ph]} \BibitemShut
  {NoStop}%
\bibitem [{\citenamefont {Artymowski}\ \emph {et~al.}(2017)\citenamefont
  {Artymowski}, \citenamefont {Lewicki},\ and\ \citenamefont
  {Wells}}]{Artymowski:2016tme}%
  \BibitemOpen
  \bibfield  {author} {\bibinfo {author} {\bibfnamefont {M.}~\bibnamefont
  {Artymowski}}, \bibinfo {author} {\bibfnamefont {M.}~\bibnamefont {Lewicki}},
  \ and\ \bibinfo {author} {\bibfnamefont {J.~D.}\ \bibnamefont {Wells}},\
  }\href {\doibase 10.1007/JHEP03(2017)066} {\bibfield  {journal} {\bibinfo
  {journal} {JHEP}\ }\textbf {\bibinfo {volume} {03}},\ \bibinfo {pages} {066}
  (\bibinfo {year} {2017})},\ \Eprint {http://arxiv.org/abs/1609.07143}
  {arXiv:1609.07143 [hep-ph]} \BibitemShut {NoStop}%
\bibitem [{\citenamefont {Redmond}\ and\ \citenamefont
  {Erickcek}(2017)}]{Redmond:2017tja}%
  \BibitemOpen
  \bibfield  {author} {\bibinfo {author} {\bibfnamefont {K.}~\bibnamefont
  {Redmond}}\ and\ \bibinfo {author} {\bibfnamefont {A.~L.}\ \bibnamefont
  {Erickcek}},\ }\href {\doibase 10.1103/PhysRevD.96.043511} {\bibfield
  {journal} {\bibinfo  {journal} {Phys. Rev.}\ }\textbf {\bibinfo {volume}
  {D96}},\ \bibinfo {pages} {043511} (\bibinfo {year} {2017})},\ \Eprint
  {http://arxiv.org/abs/1704.01056} {arXiv:1704.01056 [hep-ph]} \BibitemShut
  {NoStop}%
\bibitem [{\citenamefont {D'Eramo}\ \emph {et~al.}(2017)\citenamefont
  {D'Eramo}, \citenamefont {Fernandez},\ and\ \citenamefont
  {Profumo}}]{DEramo:2017gpl}%
  \BibitemOpen
  \bibfield  {author} {\bibinfo {author} {\bibfnamefont {F.}~\bibnamefont
  {D'Eramo}}, \bibinfo {author} {\bibfnamefont {N.}~\bibnamefont {Fernandez}},
  \ and\ \bibinfo {author} {\bibfnamefont {S.}~\bibnamefont {Profumo}},\
  }\href@noop {} {\  (\bibinfo {year} {2017})},\ \Eprint
  {http://arxiv.org/abs/1703.04793} {arXiv:1703.04793 [hep-ph]} \BibitemShut
  {NoStop}%
\bibitem [{\citenamefont {Pallis}(2006)}]{Pallis:2005bb}%
  \BibitemOpen
  \bibfield  {author} {\bibinfo {author} {\bibfnamefont {C.}~\bibnamefont
  {Pallis}},\ }\href {\doibase 10.1016/j.nuclphysb.2006.06.003} {\bibfield
  {journal} {\bibinfo  {journal} {Nucl. Phys.}\ }\textbf {\bibinfo {volume}
  {B751}},\ \bibinfo {pages} {129} (\bibinfo {year} {2006})},\ \Eprint
  {http://arxiv.org/abs/hep-ph/0510234} {arXiv:hep-ph/0510234 [hep-ph]}
  \BibitemShut {NoStop}%
\bibitem [{\citenamefont {Allahverdi}\ and\ \citenamefont
  {Drees}(2002)}]{Allahverdi:2002}%
  \BibitemOpen
  \bibfield  {author} {\bibinfo {author} {\bibfnamefont {R.}~\bibnamefont
  {Allahverdi}}\ and\ \bibinfo {author} {\bibfnamefont {M.}~\bibnamefont
  {Drees}},\ }\href {\doibase 10.1103/PhysRevLett.89.091302} {\bibfield
  {journal} {\bibinfo  {journal} {Phys. Rev. Lett.}\ }\textbf {\bibinfo
  {volume} {89}},\ \bibinfo {pages} {091302} (\bibinfo {year}
  {2002})}\BibitemShut {NoStop}%
\bibitem [{\citenamefont {Choi}\ \emph {et~al.}(2008)\citenamefont {Choi},
  \citenamefont {Kim}, \citenamefont {Lee},\ and\ \citenamefont
  {Seto}}]{Choi:2008zq}%
  \BibitemOpen
  \bibfield  {author} {\bibinfo {author} {\bibfnamefont {K.-Y.}\ \bibnamefont
  {Choi}}, \bibinfo {author} {\bibfnamefont {J.~E.}\ \bibnamefont {Kim}},
  \bibinfo {author} {\bibfnamefont {H.~M.}\ \bibnamefont {Lee}}, \ and\
  \bibinfo {author} {\bibfnamefont {O.}~\bibnamefont {Seto}},\ }\href {\doibase
  10.1103/PhysRevD.77.123501} {\bibfield  {journal} {\bibinfo  {journal} {Phys.
  Rev.}\ }\textbf {\bibinfo {volume} {D77}},\ \bibinfo {pages} {123501}
  (\bibinfo {year} {2008})},\ \Eprint {http://arxiv.org/abs/0801.0491}
  {arXiv:0801.0491 [hep-ph]} \BibitemShut {NoStop}%
\bibitem [{\citenamefont {Ackermann}\ \emph {et~al.}(2015)\citenamefont
  {Ackermann} \emph {et~al.}}]{Ackermann:2015lka}%
  \BibitemOpen
  \bibfield  {author} {\bibinfo {author} {\bibfnamefont {M.}~\bibnamefont
  {Ackermann}} \emph {et~al.} (\bibinfo {collaboration} {Fermi-LAT}),\ }\href
  {\doibase 10.1103/PhysRevD.91.122002} {\bibfield  {journal} {\bibinfo
  {journal} {Phys. Rev.}\ }\textbf {\bibinfo {volume} {D91}},\ \bibinfo {pages}
  {122002} (\bibinfo {year} {2015})},\ \Eprint
  {http://arxiv.org/abs/1506.00013} {arXiv:1506.00013 [astro-ph.HE]}
  \BibitemShut {NoStop}%
\end{thebibliography}%

\end{document}